\documentclass[aps,prx,twocolumn,showpacs,superscriptaddress]{revtex4}
\usepackage{amsmath}
\usepackage{amssymb}
\usepackage{epsfig}
\usepackage{color}
\usepackage{graphicx,amsmath}
\begin{document}

\title{Fermion condensate generates a new state of matter by making flat bands}

\author{V. R. Shaginyan}
\email{vrshag@thd.pnpi.spb.ru} \affiliation{Petersburg Nuclear
Physics Institute, Gatchina, 188300, Russia}\affiliation{Clark
Atlanta University, Atlanta, GA 30314, USA} \author{K. G.
Popov}\affiliation{Komi Science Center, Ural Division, RAS,
Syktyvkar, 167982, Russia}\author{V. A. Khodel}
\affiliation{Russian Research Centre Kurchatov Institute, Moscow,
123182, Russia} \affiliation{McDonnell Center for the Space
Sciences \& Department of Physics, Washington University, St.
Louis, MO 63130, USA}

\begin{abstract} This short review paper devoted
to 90th anniversary of S. T. Belyaev birthday. Belyaev's ideas
associated with the condensate state in Bose interacting systems
have stimulated intensive studies of the possible manifestation of
such a condensation in Fermi systems. In many Fermi systems and
compounds at zero temperature a phase transition happens that leads
to a quite specific state called fermion condensation. As a signal
of such a fermion condensation quantum phase transition (FCQPT)
serves unlimited increase of the effective mass of quasiparticles
that determines the excitation spectrum and creates flat bands. We
show that the class of Fermi liquids with the fermion condensate
forms a new state of matter. We discuss the phase diagrams and the
physical properties of systems located near that phase transition.
A common and essential feature of such systems is quasiparticles
different from those suggested by L. D. Landau, by crucial
dependence of their effective mass on temperature, external
magnetic field, pressure etc. It is demonstrated that a huge amount
of experimental data collected on different compounds suggests that
they, starting from some temperature and down, form the new state
of matter, and are governed by the fermion condensation. Our
discussion shows that the theory of fermion condensation develops
completely good description of the NFL behavior of strongly
correlated Fermi systems. Moreover, the fermion condensation can be
considered as the universal reason for the NFL behavior observed in
various HF metals, liquids, compounds with quantum spin liquids,
and quasicrystals. We show that these systems exhibit universal
scaling behavior of their thermodynamic properties. Therefore, the
quantum critical physics of different strongly correlated compounds
is universal, and emerges regardless of the underlying microscopic
details of the compounds. This uniform behavior, governed by the
universal quantum critical physics, allows us to view it as the
main characteristic of the new state of matter.

\end{abstract}



\maketitle

\section{Introduction}

\begin{figure} [! ht]
\begin{center}
\includegraphics [width=0.40\textwidth]{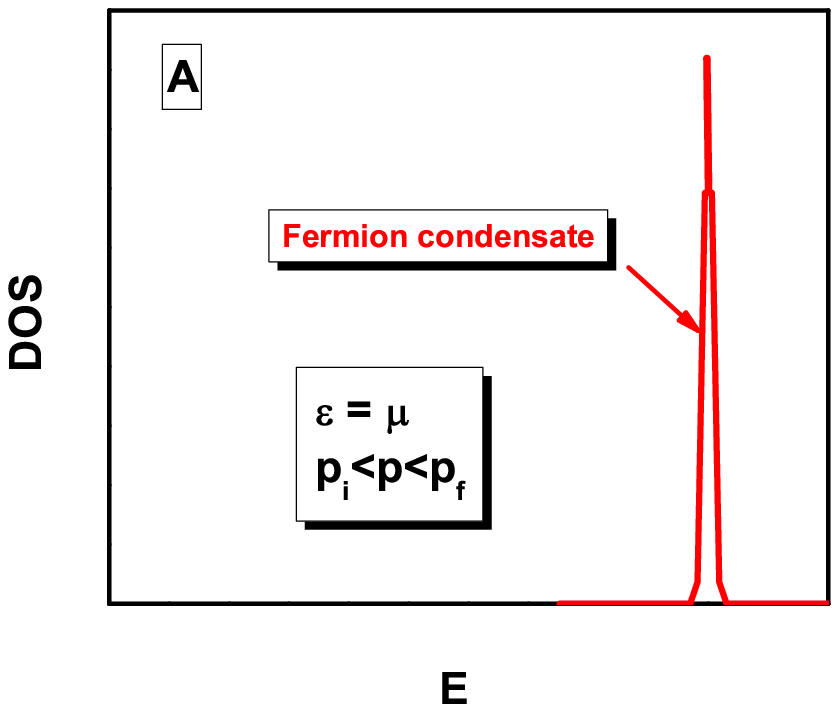}
\includegraphics [width=0.37\textwidth]{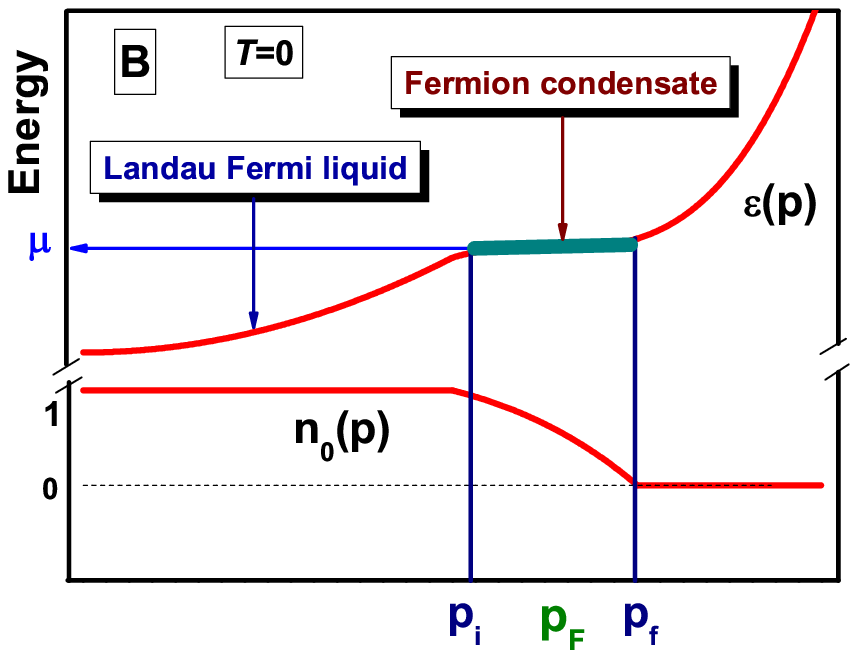}
\end{center}
\caption {(Color online). Panel {\bf A}, Schematic plot of the
density of states (DOS) of quasiparticles versus energy $E$ at the
momentum $p_1<p<p_f$ of a Fermi liquid with FC. Panel {\bf B},
Schematic plot of two-component Fermi liquid at $T=0$ with FC. The
system is separated into two parts shown by the arrows: The first
part is a Landau Fermi liquid with the quasiparticle distribution
function $n_0(p<p_i)=1$, and $n_0(p>p_f)=0$; The second one is FC
with $0<n_0(p_i<p<p_f)<1$ and the single-particle spectrum
$\varepsilon (p_i<p<p_f)=\mu$. The Fermi momentum $p_F$ satisfies
the condition $p_i<p_F<p_f$.} \label{fig0}
\end{figure}

S. T. Belyaev's contribution to theoretical physics is very
impressive. In 1958 S. T. Belyaev has published his classical works
on the theory of nonideal Bose-gas. In these works, S. T. Belyaev
demonstrates that peculiarity of Bose systems comes from a
microscopic number of particles in the condensed state with the
momentum $p=0$ \cite{bel1,bel2}. To deal with such a system, he
suggests splitting the system into two parts or subsystems, the
condensate with $p=0$ and the rest with $p>0$. It turns out that it
is the condensate that creates the "flavor" of Bose liquid,
generating its vivid properties. One may try to figure out if
physics like that of Bose systems could be represented in Fermi
systems. Belyaev's daring ideas of the two macroscopic parts in
Bose systems is adopted by a theory of fermion condensate that
permits to construct the new class of strongly correlated Fermi
liquids with the fermion condensate (FC)
\cite{ks,pr1,4,vol,volovik2}, which quasiparticle system is also
composed of two parts: One of them is represented by FC located at
the chemical potential $\mu$, and giving rise to the spiky density
of states (DOS) like that with $p=0$ of Bose system. Figure
\ref{fig0}, panel {\bf A}, shows DOS of a Fermi liquid with FC
located at the momentum $p_i<p<p_f$ and energy $\varepsilon=\mu$.
In contrast to the condensate of a Bose system occupying the $p=0$
state, quasiparticles of FC with the energy $\varepsilon=\mu$ must
be spread out over the interval $p_i<p<p_f$.

The quasiparticle distribution function $n(\bf p)$ of Fermi system
with FC is determined by the ordinary equation for a minimum of the
Landau functional $E$ \cite{ks,pr1,4}. In contrast to common
functionals of the number density $x$ \cite{hwk,wks}, the Landau
functional of the ground state energy $E$ becomes the exact
functional of the occupation numbers $n$. In case of homogeneous
system a common functional becomes a function of $x=\sum_p n({\bf
p})$, while E remains a functional, $E=E(n({\bf p}))$ \cite{pr1,4},
\begin{equation} \frac{\delta E(n({\bf p}))}{\delta
n({\bf p})}=\varepsilon({\bf p})=\mu;\,\, {\rm if:\,} 0<n({\bf
p})<1. \label{FCM}\end{equation} Equation \eqref{FCM} represents an
ordinary one to search the minimum of functional $E$. In the case
of Bose system the equation $\delta E/\delta n(p)=\mu$ describes a
common instance. In the case of Fermi systems such an equation,
generally speaking, were not correct. Thus, it is the binding
constraint $0<n({\bf p})<1$, taking place over some region
$p_i<p<p_f$, that makes Eq. \eqref{FCM} applicable for Fermi
systems. Because of the binding constraint, Fermi quasiparticles in
the region $p_i<p<p_f$ can behave as Bose one, occupying the same
energy level $\varepsilon=\mu$, and Eq. \eqref{FCM} yields the
quasiparticle distribution function $n_0({\bf p})$ that minimizes
the ground-state energy $E$. A possible solution $n_0({\bf p})$ of
Eq. (\ref{FCM}) and the corresponding single-particle spectrum
$\varepsilon({\bf p})$ are depicted in Fig. \ref{fig0}, panel {\bf
B}. As seen from the panel {\bf B}, $n_0({\bf p})$ differs from the
step function in the interval $p_i<p<p_f$, where $0<n_0({\bf
p})<1$, and coincides with the step function outside this interval.
Thus, the Fermi surface at $p=p_F$ transforms into the Fermi volume
at $p_i\leq p\leq p_f$ suggesting that the band is absolutely
``flat'' within this interval, giving rise to the spiky DOS. The
existence of such flat bands formed by inter-particle interaction
has been predicted for the first time in Ref. \cite{ks}.
Quasiparticles with momenta within the interval $(p_i<p<p_f)$ have
the same single-particle energies equal to the chemical potential
$\mu$ and form FC, while the distribution $n_0({\bf p})$ describes
the new state of the Fermi liquid with FC, and the Fermi system is
split up into to parts: a Landau Fermi liquid (LFL) and the FC
part, as it is shown in Fig. \ref{fig0}, panel {\bf B}
\cite{ks,pr1,vol,volovik2,4}.

In contrast to the Landau, marginal, or Luttinger Fermi liquids,
which exhibit the same topological structure of the Green's
function, in systems with FC, where the Fermi surface spreads into
the Fermi volume, the Green's function belongs to a different
topological class. The topological class of the Fermi liquid is
characterized by the invariant \cite{vol,volovik2}
\begin{equation} N=tr\oint_C\frac{dl}{2\pi i}G(i\omega,{\bf p})
\partial_lG^{-1}(i\omega,{\bf p})\label{FLVOL},\end{equation}
where ``tr'' denotes the trace over the spin indices of the Green's
function and the integral is taken along an arbitrary contour $C$
encircling the singularity of the Green's function. The invariant
$N$ in \eqref{FLVOL} takes integer values even when the singularity
is not of the pole type, cannot vary continuously, and is conserved
in a transition from the Landau Fermi liquid to marginal liquids
and under small perturbations of the Green's function. As shown by
Volovik \cite{vol,volovik2}, the situation is quite different for
systems with FC, where the invariant $N$ becomes a half-integer and
the system with FC transforms into an entirely new class of Fermi
liquids with its own topological structure.

In contrast to Bose liquid, which entropy $S\to0$ at temperature
$T\to0$, a Fermi liquid with FC possesses finite entropy $S_0$ at
zero temperature \cite{4,aplh}. Indeed, as it is seen from Fig.
\ref{fig0}, panel {\bf B}, at $T=0$, the ground state of a system
with a flat band is degenerate, and the occupation numbers
$n_0({\bf p})$ of single-particle states belonging to the flat band
are continuous functions of momentum ${\bf p}$, in contrast to
discrete standard LFL values 0 and 1. Such behavior of the
occupation numbers leads to a $T$-independent entropy term
$S_0=S(T\to0,n=n_0)$ with the entropy given by
\begin{equation}
S(n)=-\sum_{\,{\bf p}} [n({\bf p})\ln n({\bf p})+(1-n({\bf
p}))\ln(1- n({\bf p}))].\label{S0}
\end{equation}
Since the state of a system with FC is highly degenerate, FC serves
as a stimulator of phase transitions that could lift the degeneracy
of the spectrum and make $S_0$ vanish in accordance with the Nernst
theorem. For instance, FC can excite the formation of spin density
waves, antiferromagnetic state and ferromagnetic state etc., thus
strongly stimulating the competition between phase transitions
eliminating the degeneracy. The presence of FC facilitates a
transition to the superconducting state, because the both phases
have the same order parameter \cite{ks,4}. Thus, in contrast to
Bose systems with Bose condensate, entropy of which at lowering
temperatures $S\to0$, the $S_0$ peculiarity of Fermi systems with
FC incites to the emerging of great diversity of states. Being
generated by the same driving motive - $S_0$ - these, as we shall
see, exhibit a universal behavior, and form a new state of matter
demonstrated by many compounds.

In this paper we briefly review the theory of FC that permits to
describe a tremendously broad variety of experimental results in
different systems. We assume that these systems are located near
the fermion condensation quantum phase transition (FCQPT), leading
to the emergence of FC \cite{4,ks}. The rest of the paper is
organized as follows. In Section 2, we examine a scaling behavior
of the effective mass and heavy fermion (HF) metals based on the
extended quasiparticle paradigm that is employed to renovate the
Landau quasiparticle paradigm. In Section 3, we construct phase
diagrams of Fermi systems with FC, and compare these with the
experimental ones, and show that FC leads to a new state of matter.
In Section 4, we apply the FC theory to describe the thermodynamic
properties of strongly correlated Fermi systems represented  by
compounds holding a quantum spin liquid. Section 5 is devoted to
quasicrystals. Section 6 summaries the main results.

\section{Extended quasiparticle paradigm}

\begin{figure} [! ht]
\begin{center}
\includegraphics [width=0.47\textwidth]{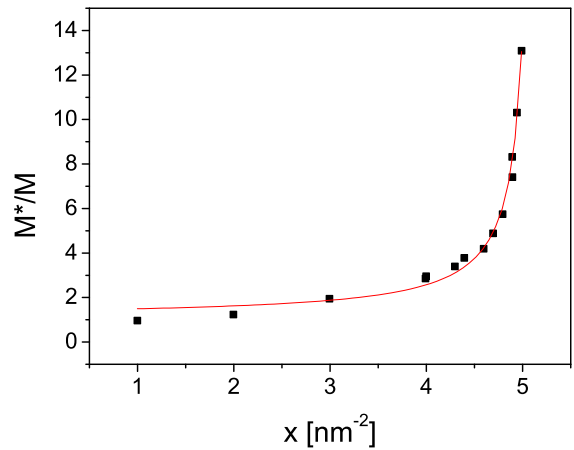}
\end{center}
\caption {(Color online). The ratio $M ^*/M$ in two-dimensional
$^3$He as a function of the number density $x$ of the liquid,
obtained from heat capacity and magnetization measurements. The
experimental data are marked by black squares \cite{8}, and the
solid line represents the function given by Eq. \eqref{FL7}.}
\label {Fig6}
\end{figure}

Upon using the Landau functional, one can obtain the well-known
Landau equation for the effective mass \cite{4,1_1,1_2,lanl}
\begin{equation}\label{FLL} \frac{1}{M^*} =
\frac{1}{M}+\int \frac{{\bf p}_F{\bf p_1}}{p_F^3} F({\bf p_F},{\bf
p}_1,n)\frac{\partial n(\bf p_1)}{\partial {\bf p_1}} \frac{d{\bf
p}_1}{(2\pi)^3}.
\end{equation}
Here, $F$ is the Landau interaction. For simplicity, we omit the
spin dependencies. To calculate $M^*$ as a function of $T$, we
construct the free energy $F=E-TS$, where $S$ is given by
\eqref{S0}. Minimizing $F$ with respect to $n({\bf p})$, we arrive
at the Fermi-Dirac distribution
\begin{equation} n_{\pm}({\bf
p},T)= \left\{1+\exp\left[\frac{(\varepsilon({\bf p},T)\pm
B\mu_B-\mu)} {T}\right]\right\}^{-1}.\label{FL4}
\end{equation}
Here $\mu$ is the chemical potential, $B$ is an external magnetic
field, and $\mu_B$ is the Bohr magneton. The term $\pm B\mu_B$
entering the right hand side of Eq. \eqref{FL4} describes Zeeman
splitting. Eq. \eqref{FLL} is exact, and allows us to calculate the
behavior of $M^*$ which now becomes a function of temperature $T$,
external magnetic field $B$, number density $x$ and pressure $P$,
rather than a constant. It is this feature of $M^*$ that forms both
the scaling and the non-Fermi liquid (NFL) behavior observed in
measurements on strongly correlated Fermi systems \cite{4,UBH,UFN}.
In case of finite $M^*$ and at $T=0$ the distribution function
$n({\bf p},T=0)$ becomes the theta-function $\theta (p_F-p)$, as it
follows from \eqref{FL4}, and Eq. (\ref{FLL}) yields the well-known
result
\begin{equation}\label{M*}
\frac{M^*}{M}=\frac{1}{1-F^1/3}.
\end{equation}
where $F^1=N_0f^1$, $f^1(p_F,p_F)$ is the $p$-wave component of the
Landau interaction, $M$ is the bare mass of particles of the liquid
in question, and $N_0=Mp_F/(2\pi^2)$ is the DOS of a free Fermi
gas. Because $x=p_F^3/3\pi^2$ in the Landau Fermi-liquid theory,
the Landau interaction can be written as $F^1(p_F,p_F)=F^1(x)$. At
a certain critical point $x=x_{FC}$ the denominator $(1-F^1(x)/3)$
tends to zero, $(1-F^1(x)/3)\propto(x-x_{FC})+a(x-x_{FC})^2 +
...\to 0$, and we find that
\begin{equation}
\frac{M^*(x)}{M}\simeq a_1+\frac{a_2}{x-x_{FC}}.
\label{FL7}\end{equation} where $a_1$ and $a_2$ are constants and
$M^*(x\to x_{FC})\to\infty$. As a result, at $T\to0$ and $x\to
x_{FC}$ the system undergoes a quantum phase transition
\cite{4,UFN}, represented by FCQPT. We note that the divergence of
the effective mass given by Eq. \eqref{FL7} does preserve the
Pomeranchuk stability conditions, for $F^1$ is positive
\cite{pom,noz}. The behavior of $M^*(x)$ given by (\ref{FL7}) is in
good agreement with the results of experiments on $^3$He \cite{8}.
The divergence of the effective mass  $M^*(x)$ observed in
measurements on two-dimensional $^3$He \cite{8} is illustrated in
Fig. \ref{Fig6}. It is seen that calculations based on \eqref{FL7}
are in good agreement with the experimental data.

It is instructive to briefly explore the behavior of $M^*$ in order
to capture its universal behavior at FCQPT. Let us write the
quasiparticle distribution function as $n_1({\bf p})=n({\bf
p},T,B)-n({\bf p})$, with $n({\bf p})$ being the step function, and
Eq. \eqref{FLL} then becomes
\begin{equation}
\frac{1}{M^*(T,B)}=\frac{1}{M^*}+\int\frac{{\bf p}_F{\bf
p_1}}{p_F^3}F({\bf p_F},{\bf p}_1)\frac{\partial n_1(p_1,T,B)}
{\partial p_1}\frac{d{\bf p}_1}{(2\pi)^3}. \label{LF1}
\end{equation}
At FCQPT, that is at $x\to x_{FC}$, the effective mass $M^*$
diverges and Eq. \eqref{LF1} becomes homogeneous, determining
$M^*(T,B)$ as a universal function of temperature and magnetic
field. The only role of $F$ is to drive the system to FCQPT and the
solutions of Eq. \eqref{LF1} are represented by some universal
function of variables $T$, $B$, $x$. In that case $M^*$ strongly
depends on the same variables. In contrast to the Landau
quasiparticle paradigm assuming the constancy of the effective
mass, the extended quasiparticle paradigm is to be introduced
\cite{4}. The main point here is that the well-defined
quasiparticles determine as before the thermodynamic, relaxation
and transport properties of strongly correlated Fermi-systems,
while $M^*$ becomes a function of $T$, $B$, $x$, etc.

\subsection{Scaling behavior of both the effective mass and HF
metals}

A deeper insight into the behavior of $M^*(T,B)$ can be achieved
using some "internal" scales. Namely, near FCQPT the solutions of
Eq. \eqref{LF1} exhibit a behavior so that $M^*(T,B)$ reaches its
maximum value $M^*_M$ at some temperature $T_{M}\propto B$
\cite{4}. It is convenient to introduce the internal scales $M^*_M$
and $T_{M}$ to measure the effective mass and temperature. Thus, we
divide the effective mass $M^*$ and the temperature $T$ by the
values, $M^*_M$ and $T_{M}$, respectively. This generates the
normalized effective mass $M^*_N=M^*/M^*_M$ and the normalized
temperature $T_N=T/T_{M}$.

\begin{figure} [! ht]
\begin{center}
\vspace*{-0.2cm}
\includegraphics [width=0.47\textwidth]{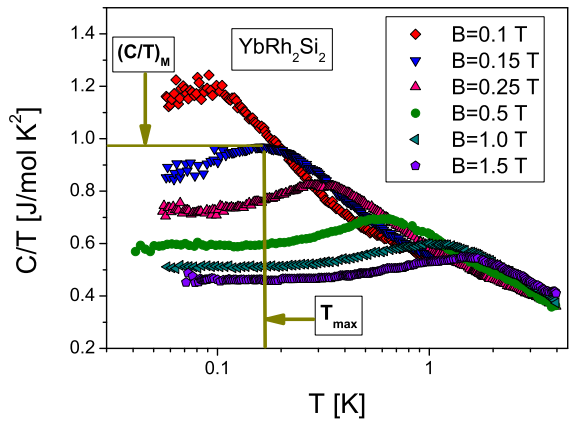}
\end{center}
\vspace*{-0.3cm} \caption{(Color online). Electronic specific heat
of $\rm YbRh_2Si_2$, $C/T$, versus temperature $T$ as a function of
magnetic field $B$ \cite{oesb} shown in the legend. The
illustrative values of $(C/T)_M\propto M^*_M$ and $T_{M}$ at
$B=0.15$ T are shown.}\label{FIG2}
\end{figure}
As illustration to the above consideration, we analyze the specific
heat $C/T\propto M^*$ of the HF metal $\rm YbRh_2Si_2$ \cite{oesb}.
Under the application of magnetic field $B$, the specific heat
exhibits a behavior that is described by a function of both $T$ and
$B$.
\begin{figure} [! ht]
\begin{center}
\vspace*{-0.2cm}
\includegraphics [width=0.47\textwidth]{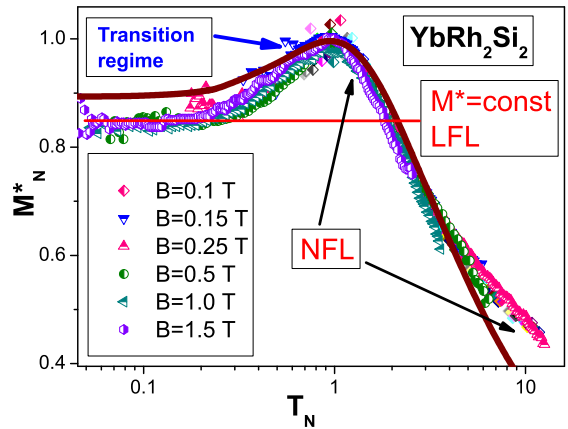}
\end{center}
\vspace*{-0.3cm} \caption{(Color online). The normalized effective
mass $M^*_N$ versus normalized temperature $T_N$.  $M^*_N$ is
extracted from the measurements of the specific heat $C/T$ on $\rm
YbRh_2Si_2$ in magnetic fields $B$ \cite{oesb} listed in the
legend. Constant effective mass $M^*_L$ inherent in a normal Landau
Fermi liquid (LFL) is depicted by the solid line. The transition
region is shown by the arrow and the NFL behavior is depicted by
two arrows. Our calculation based on Eq. \eqref{FLL} is displayed
by the solid curve.}\label{FIG3}
\end{figure}
As seen from Fig. \ref{FIG2}, a maximum structure $(C/T)_M$ in
$C/T\propto M^*_M$ at temperature $T_{M}$ appears under the
application of magnetic field $B$. $T_{M}$ shifts to higher $T$ and
$C/T\propto M^*_M$ diminishes as $B$ is increased. The value $C/T$
is saturated towards lower temperatures, decreasing at elevated
magnetic field. To obtain the normalized effective mass $M^*_N$, we
use $(C/T)_M$ and $T_{M}$ as "internal" scales: The maximum
structure $(C/T)_M$ was used to normalize $C/T$, and $T$ was
normalized by $T_{M}$. In Fig. \ref{FIG3}, the obtained
$M^*_N=(C/T)/(C/T)_M=M^*/M^*_M$, as a function of normalized
temperature $T_N=T/T_{M}$, is shown by geometrical figures. It is
seen that the Landau Fermi liquid (LFL) state and NFL one are
separated by the transition regime at which $M^*_N$ reaches its
maximum value. Figure \ref{FIG3} reveals the scaling behavior of
the normalized experimental curves - the curves at different
magnetic fields $B$ merge into a single one in terms of the
normalized variable $T/T_M$. Our calculations of the normalized
effective mass $M^*_N(T_N)$, shown by the solid line and based on
Eq. \eqref{LF1}, are in good agreement with the experimental facts
\cite{4,dft373}. Near FCQPT the normalized solution of Eq.
\eqref{FLL} $M^*_N(T_N)$ can be well approximated by a simple
universal interpolating function \cite{4}. The interpolation occurs
between the LFL and NFL regimes and represents the universal
scaling behavior of $M^*_N$ \cite{4,dft373}
\begin{equation}M^*_N(T_N)\approx c_0\frac{1+c_1T_N^2}{1+c_2T_N^{n}}.
\label{UN2}
\end{equation}
Here, $c_0=(1+c_2)/(1+c_1)$, $c_1$, $c_2$ are fitting parameters;
and the exponent $n=8/3$ if the Landau interaction is an analytical
function otherwise $n=5/2$ \cite{4,14}. It follows from
Eq.~\eqref{UN2} that
\begin{equation}
\label{TMB} T_M\simeq a_1\mu_BB,
\end{equation}
where $a_1$ is a dimensionless factor.

Several remarks concerning the applicability of Eqs. \eqref{FLL}
and \eqref{UN2} to systems with violated translational invariance
are in order. We study the universal behavior of HF metals, quantum
spin liquids, and quasicrystals at low temperatures using the model
of a homogeneous HF liquid \cite{4}. The model is applicable
because we consider the scaling behavior exhibited by the
thermodynamic properties of these materials at low temperatures, a
behavior related to the scaling of quantities such as the effective
mass $M^*$, the heat capacity $C/T\propto M^*$, the magnetic
susceptibility $\chi\propto M^*$, etc. The behavior of $M^*_N(T_N)$
that characterizes them are determined by momentum transfers that
are small compared to momenta of the order of the reciprocal
lattice length. The high momentum contributions can therefore be
ignored by substituting the lattice for the jelly model. While the
values of the scales like the maximum $M^*_M(B_0)$ of the effective
mass, measured at some field $B=B_0$, and $T_M$ at which $M^*_M$
takes place are determined by a wide range of momenta. Thus, these
scales are controlled by the specific properties of the system
under consideration. It follows from our consideration that the
scaled thermodynamic properties of different strongly correlated
Fermi systems can be described by universal function \eqref{UN2}
determining $M^*_N(T_N)$. It is seen from Fig. \ref{FIG3}, and
demonstrated by a huge amount of experimental data collected on
strongly correlated Fermi systems that this observation is in good
agreement with experimental facts \cite{4}.

\section{Phase diagrams of strongly correlated Fermi systems}

At $T=0$, a quantum phase transition is driven by a nonthermal
control parameter, e.g. the number density $x$, magnetic field $B$,
pressure $P$. At the quantum critical point (QCP) related to FCQPT
and taking place at $x=x_{FC}$, the effective mass $M^*$ diverges.
We note that there are different kinds of instabilities of normal
Fermi liquids connected with several perturbations of initial
quasiparticle spectrum $\varepsilon(p)$ and occupation numbers
$n(p)$, associated with the emergence of a multi-connected Fermi
surface, see e.g. \cite{4,khodb,asp,zvbld}. Depending on the
parameters and analytical properties of the Landau interaction,
such instabilities lead to several possible types of restructuring
of the initial Fermi liquid ground state. This restructuring
generates topologically distinct phases. One of them is the FC,
another one belongs to a class of topological phase transitions,
where the sequence of rectangles $n(p)=0$ and $n(p)=1$ is realized
at $T=0$. In fact, at elevated temperatures the systems located at
these transitions exhibit behavior typical to those located at
FCQPT \cite{4}. Therefore, we do not consider the specific
properties of these topological transitions, and focus on the
behavior of systems located near FCQPT.

\begin{figure} [! ht]
\begin{center}
\vspace*{-0.2cm}
\includegraphics [width=0.47\textwidth]{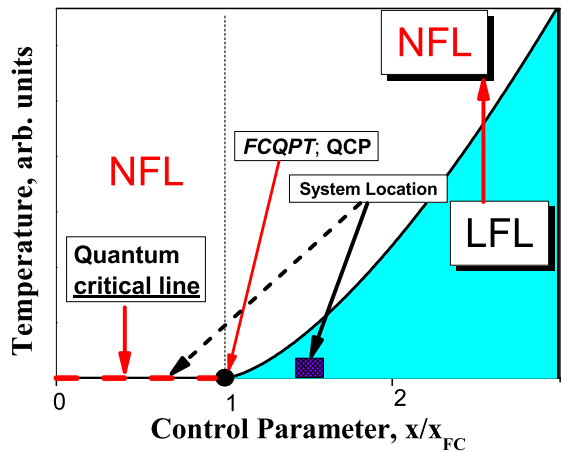}
\end{center}
\vspace*{-0.3cm} \caption{(Color online). Schematic $T-x$ phase
diagram of system with FC. The number density $x$ is taken as the
control parameter and depicted as $x/x_{FC}$. At $x/x_{FC}>1$ and
sufficiently low temperatures, the system is in the LFL state as
shown by the shadow area. This location of the system is depicted
by both the solid square and the arrow. The vertical arrow
illustrates the system moving in the LFL-NFL direction along $T$ at
fixed control parameter.  At $x/x_{FC}<1$ the system is shifted
beyond the QCP, and is at the quantum critical line depicted by the
dashed line and shown by the vertical arrow. This location of the
system is shown by the short dash arrow. At any finite low
temperatures $T>0$ the system possesses finite entropy $S_0$ and
exhibits the NFL behavior.}\label{fig03}
\end{figure}

Beyond FCQPT, the system forms FC that leads to the emergence of a
flat band protected by topological invariants \cite{vol,volovik2}.
The $T-x$ schematic phase diagram of the system which is driven to
the FC state by variation of the number density $x$ is reported in
Fig. \ref{fig03}. Upon approaching the critical density $x_{\rm
FC}$ the system remains in the LFL region at sufficiently low
temperatures, as it is shown by the shadow area. The temperature
range of the shadow area shrinks as the system approaches QCP, and
$M^*(x\to x_{FC})$ diverges as it follows from Eq. \eqref{FL7}. At
this QCP shown by the arrow in Fig. \ref{fig03}, the system
demonstrates the NFL behavior down to the lowest temperatures.
Beyond the critical point at finite temperatures the behavior
remains the NFL, and is determined by the temperature-independent
entropy $S_0$ \cite{4,yakov}. In that case at $T\to 0$, the system
again demonstrates the NFL behavior, and approaches a quantum
critical line (QCL) (shown by the vertical arrow and the dashed
line in Fig. \ref{fig03}) rather than a quantum critical point.
Upon reaching the quantum critical line from the above at $T\to0$
the system undergoes the first order quantum phase transition,
making $S_0$ vanish. As it is seen from Fig. \ref{fig03}, at rising
temperatures the system located before QCP does not undergo a phase
transition, and transits from the NFL behavior to the LFL one. At
finite temperatures there is no boundary (or phase transition)
between the states of the system located before or behind QCP,
shown by the arrows. Therefore, at elevated temperatures the
properties of systems with $x/x_{\rm FC}<1$ or with $x/x_{FC}>1$
become indistinguishable.

As seen from Fig. \ref{fig03}, the location of the system is
controlled by the number density $x$. At rising $x$, $x/x_{FC}>1$,
the system is located before FCQPT, and demonstrates the LFL
behavior at low temperatures. We speculate that such a state can be
induced by the application of positive pressure, including positive
chemical pressure. On the other hand, at diminishing $x$,
$x/x_{FC}<1$, the system is shifted beyond FCQPT, and is at the
quantum critical line depicted by the dashed line. In that case the
system demonstrates the NFL behavior at any finite temperatures. We
assume that such a state can be induced by the application of
negative pressure, including negative chemical pressure. At low
temperatures and above the critical line, the system has the finite
entropy $S_0$ and its NFL state is strongly degenerate. The
degeneracy stimulates the emergence of different phase transitions,
lifting it and removing the entropy $S_0$. The NFL state can be
captured by other states such as superconducting, for example, by
the superconducting state (SC) in $\rm CeCoIn_5$, or by
antiferromagnetic (AF) state, e.g. the AF in $\rm YbRh_2Si_2$
\cite{4,dft373,yakov}. The diversity of phase transitions occurring
at low temperatures is one of the most spectacular features of the
physics of many HF metals and strongly correlated compounds. Within
the scenario of ordinary quantum phase transitions, it is hard to
understand why these transitions are so different from one another
and their critical temperatures are so extremely small. However,
such diversity is endemic to systems with a FC, since the FC state
must be altered as $T\to 0$, so that the excess entropy $S_0$ is
shed before zero temperature is reached. At finite temperatures
this takes place by means of some phase transitions which can
compete, shedding the excess entropy \cite{4,khodb}.

\begin{figure}[!ht]
\begin{center}
\includegraphics [width=0.48\textwidth]{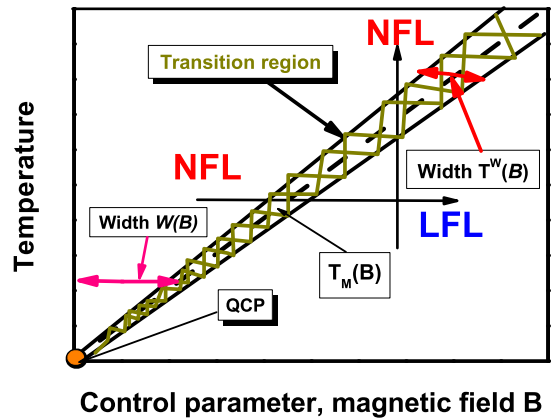}
\end{center}
\caption{(Color online). Schematic $T-B$ phase diagram of HF liquid
with magnetic field as the control parameter. The vertical and
horizontal arrows show LFL-NFL and NFL-LFL transitions at fixed $B$
and $T$, respectively. At $B=0$ the system is in its NFL state,
having a flat band, and demonstrates NFL behavior down to $T\to0$.
The hatched area separates the NFL phase and the weakly polarized
LFL phase and represents the transition region. The dashed line in
the hatched area represents the function $T_M(B)\simeq T^*_{FC}$
given by Eq.~\eqref{TMB}. The functions $W(B)\propto T$ and
$T^W(B)\propto T$ shown by two-headed arrows define the width of
the NFL state and the transition area, respectively. The QCP
located at the origin and indicated by the arrow denotes the
critical point at which the effective mass $M^*$ diverges and both
$W(B)$ and $T^W(B)$ tend to zero.}\label{fig1}
\end{figure}
The schematic $T-B$ phase diagram of a HF liquid is depicted in
Fig.~\ref{fig1}, with the magnetic field $B$ serving as the control
parameter.  At $B=0$, the HF liquid acquires a flat band
corresponding to a strongly degenerate state. The NFL regime reigns
at elevated temperatures and fixed magnetic field.  With increasing
$B$, the system is driven from the NFL region to the LFL domain. As
shown in Fig.~\ref{fig1}, the system moves from the NFL regime to
the LFL regime along a horizontal arrow, and from the LFL to NFL
along a vertical arrow.  The magnetic-field-tuned QCP is indicated
by the arrow and located at the origin of the phase diagram, since
application of any magnetic field destroys the flat band and shifts
the system into the LFL state \cite{4,shag,mig100,takah,geg}. The
hatched area denoting the transition region separates the NFL state
from the weakly polarized LFL state and contains the dashed line
tracing the transition region, $T_M(B)\simeq T^*_{FC}$. Referring
to Eq.~\eqref{TMB}, this line is defined by the function
$T^*_{FC}\propto \mu_BB$, and the width $W(B)$ of the NFL state is
seen to be proportional $T$. In the same way, it can be shown that
the width $T^W(B)$ of the transition region is also proportional to
$T$.

\begin{figure}[!ht]
\begin{center}
\includegraphics [width=0.48\textwidth]{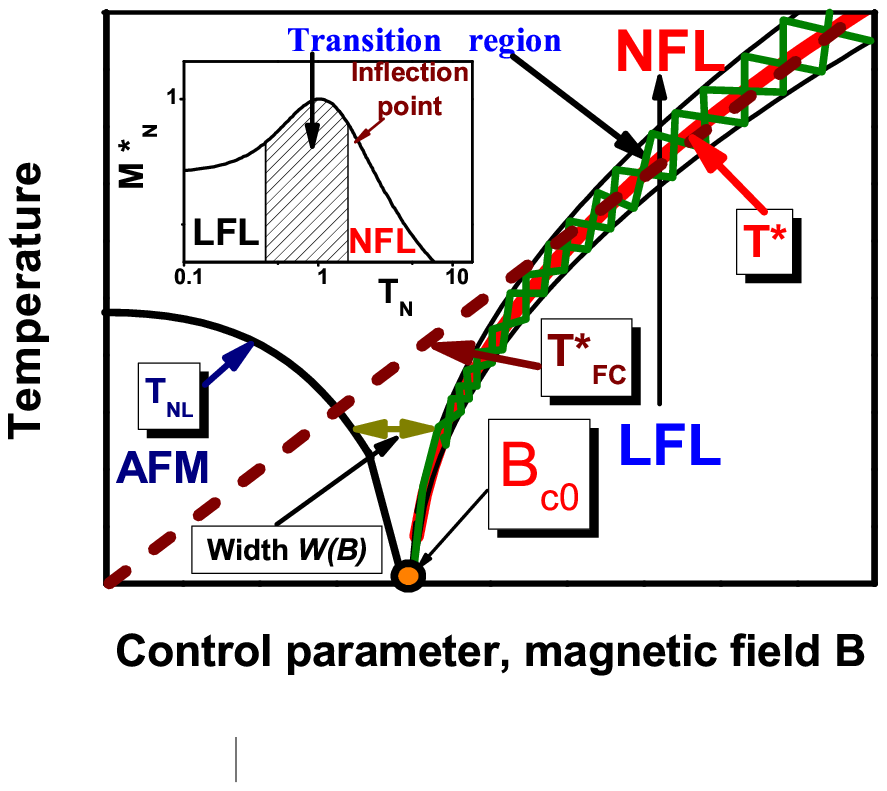}
\vspace*{-1.5cm}
\end{center}
\caption{(Color online). Schematic $T-B$ phase diagram of a HF
metal with magnetic field as the control parameter. The AF phase
boundary line is shown by the arrow and depicted by the solid
curve, representing the N\'eel temperature $T_{NL}$. The vertical
and horizontal arrows show LFL-NFL and NFL-LFL transitions at fixed
$B$ and $T$, respectively. The hatched area separates the NFL phase
and the weakly polarized LFL phase and represents the transition
area. The solid curve in the hatched area represents the transition
temperature $T^*$. The solid short dash line $T^*_{FC}(B)\propto
B\mu_B$ depicts the transition temperature provided that the AF
state were absent. The functions $W(B)\propto T$ shown by
two-headed arrows defines the total width of both the NFL state and
the transition area. The inset shows a schematic plot of the
normalized effective mass versus the normalized temperature. The
transition regime, where $M^*_N$ reaches its maximum value at
$T_N=T/T_M=1$, is shown as the hatched area in both the main panel
and the inset.  Arrows indicate the transition region and the
inflection point $T_{inf}$ in the $M^*_N$ plot.}\label{fig04}
\end{figure}

\begin{figure} [! ht]
\begin{center}
\vspace*{-0.2cm}
\includegraphics [width=0.47\textwidth]{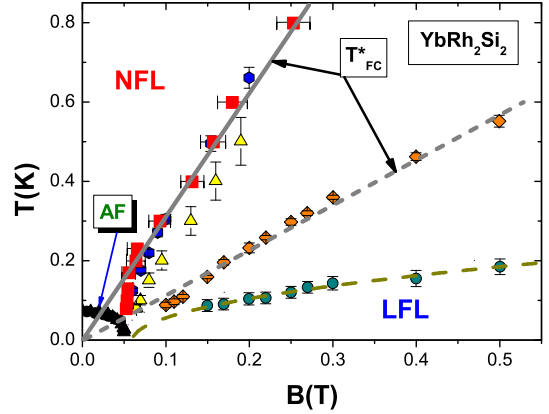}
\end{center}
\vspace*{-0.3cm} \caption{(Color online). Temperature versus
magnetic field $T-B$ phase diagram for $\rm YbRh_2Si_2$. Solid
circles represent the boundary between AF and NFL states. The curve
is represented by the function $\sqrt{B-B_{c0}}$ \cite{4}. The
diamonds along the short-dash curve denote the transition area
between the NFL and LFL regime \cite{steg1,geg}. The short-dash
line is approximated by $T^*_{FC}\propto B\mu_B$, see Fig.
\ref{fig04}. Diamonds marking $T^*$ along this dash line signify
the maxima $T_M$ of $C/T$ \cite{oesb} shown in Fig. \ref{FIG2}. The
transition temperature $T^*$ as determined from magnetostriction
(solid squares), longitudinal magnetoresistivity (triangles), and
susceptibility (solid circles) \cite{steg1}. The solid line depicts
the linear fit to $T^*$.}\label{f02}
\end{figure}
Now we construct the $T-B$ schematic phase diagram of a HF metal
like $\rm YbRh_2Si_2$ shown in Fig. \ref{fig04} \cite{4,jrho}. We
start with observing that Eq. \eqref{UN2} reveals the scaling
behavior of the normalized effective mass $M^*_N(T_N)$: Values of
the effective mass $M^*(T,B)$ at different magnetic fields $B$
merge into a single mass value $M^*_N$ in terms of the normalized
variable $T_N=T/T_M$. The inset in Fig. \ref{fig04} demonstrates
that scaling behavior of the normalized effective mass $M^*_N$
versus the normalized temperature $T_N$. The LFL phase prevails at
$T\ll T_M$, followed by the $T^{-\beta}$ regime at $T \gtrsim T_M$,
with $\beta=2/3$ or $\beta=1/2$ as it follows from Eq. \eqref{UN2}.
The latter phase is designated as NFL due to the strong dependence
of the effective mass on temperature. The temperature region
$T\simeq T_M$ encompasses the transition between the LFL regime
with almost constant effective mass and the NFL behavior. Thus
$T^*\sim T_M(B)$ identifies the transition region featuring a
crossover between LFL and NFL regimes. The transition (crossover)
temperature $T^*(B)$ is not actually the temperature of a phase
transition.  Its specification is necessarily ambiguous, depending
as it does on the criteria invoked for determination of the
crossover point. As usual, the temperature $T^*(B)$ is extracted
from the field dependence of charge transport, for example from the
resistivity $\rho(T)$ given by
\begin{equation}
\rho(T)=\rho_0+AT^{\alpha_R},\label{res}
\end{equation}
where $\rho_0$ is the residual resistivity and $A$ is a
$T$-independent coefficient. The term $\rho_0$ is ordinarily
attributed to impurity scattering. The LFL state is characterized
by the $T^{\alpha_R}$ dependence of the resistivity with index
$\alpha_R=2$. The crossover (through the transition regime shown as
the hatched area in both Fig.~\ref{fig04} and its inset) takes
place at temperatures where the resistance starts to deviate from
LFL behavior, with the exponent $\alpha_R$ shifting from 2 into the
range  $1<\alpha_R<2$. When constructing the phase diagram depicted
in Fig.~\ref{fig04}, we assume that AF order wins the competition,
destroying the $S_0$ term at low temperatures. At $B=B_{c0}$, the
HF liquid acquires a flat band corresponding to a strongly
degenerate state. Here, $B_{c0}$ is a critical magnetic field, such
that at $T\to0$ the application of magnetic field $B\gtrsim B_{c0}$
destroys the AF state restoring the paramagnetic state with the LFL
behavior. In some cases as in the HF metal $\rm CeRu_2Si_2$,
$B_{c0}=0$, see e.g. \cite{takah}, while in $\rm YbRh_2Si_2$,
$B_{c0}\simeq 0.06$ T \cite{geg}. Obviously, $B_{c0}$ is defined by
the specific system properties, therefore we consider it as a
parameter. The NFL regime reigns at elevated temperatures and fixed
magnetic field. With increasing $B$, the system is driven from the
NFL region to the LFL domain. As shown in Figs. \ref{fig1} and
\ref{fig04}, the system moves from the NFL regime to the LFL regime
along the corresponding arrows. The magnetic-field-tuned QCP is
indicated by the arrow and located at $B=B_{c0}$. The hatched area
denotes the transition region, and separates the NFL state from the
weakly polarized LFL state. It contains both the dashed line
tracing $T^*_{FC}(B)$ and the solid curve $T^*(B)$. Referring to
Eq. \eqref{UN2}, the latter is defined by the function
$T^*\propto\mu_BB$ and merges with $T^*_{FC}(B)$ at relatively high
temperatures, and $T^*\propto\mu_B(B-B_{c0})$ at lower $T\sim
T_{NL}$, with $T_{NL}(B)$ being the N\'eel temperature. As seen
from Eq. \eqref{UN2}, both the width $W(B)$ of the NFL state and
the width of the transition region are proportional $T$
\cite{jrho}. The AF phase boundary line is shown by the arrow and
depicted by the solid curve. As it was mentioned above, the dashed
line $T^*_{FC}(B)\propto B\mu_B$ represents the transition
temperature provided that the AF state were absent. In that case
the FC state is destroyed by any weak magnetic field $B\to0$ at
$T\to0$ and the dashed line $T^*_{FC}$ crosses the origin of
coordinates, as it is displayed in Figs. \ref{fig1} and
\ref{fig04}. At $T\gtrsim T_{NL}(B=0)$ the transition temperature
$T^*_{FC}(B)$ coincides with $T^*(B)$ shown by the solid curve,
since the properties of the system are given by its local free
energy, describing the paramagnetic state of the system. One might
say that the system "does not remember" the AF state, emerging at
lower temperatures. This observation is in good agreement with
experimental facts collected on the HF metal $\rm YbRh_2Si_2$.
These experimental facts are summarized in the phase diagrams of
Fig. \ref{f02}: At relatively high temperatures $T\gtrsim
T_{NF}(B=0)$ the transition temperature $T^*$, obtained in
measurements on $\rm YbRh_2Si_2$ \cite{steg1,geg}, is well
approximated by the straight lines representing $T^*_{FC}$. It is
seen from Fig. \ref{f02}, that the slope of the short dash line
(representing the maxima of the specific heat $C/T$) is different
from the slope of the solid line (representing maxima of the
susceptibility $\chi(T)$). Such a behavior is determined by the
fact that the maxima of $C/T$ and $\chi(T)$ are given by the
different relations, determining the inflection points of the
entropy: $\partial^2S/\partial T^2=0$ and $\partial^2S/\partial
B^2=0$, correspondingly. The theory of FC shows that the inflection
points exist, provided that the system is located near FCQPT
\cite{4,jetp90}.

\begin{figure} [! ht]
\begin{center}
\includegraphics [width=0.42\textwidth]{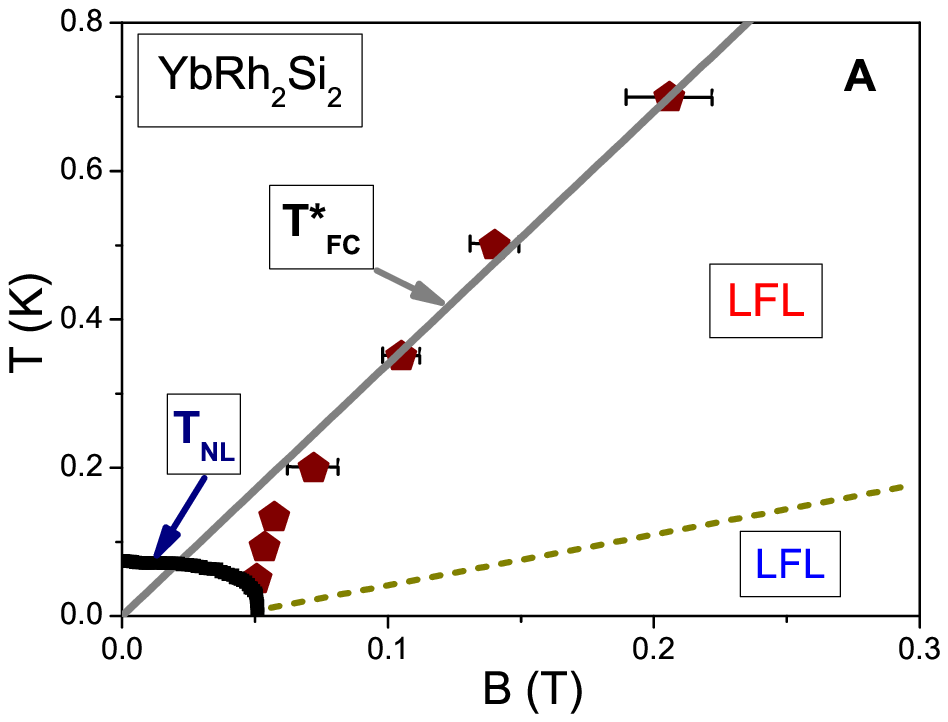}
\includegraphics [width=0.42\textwidth]{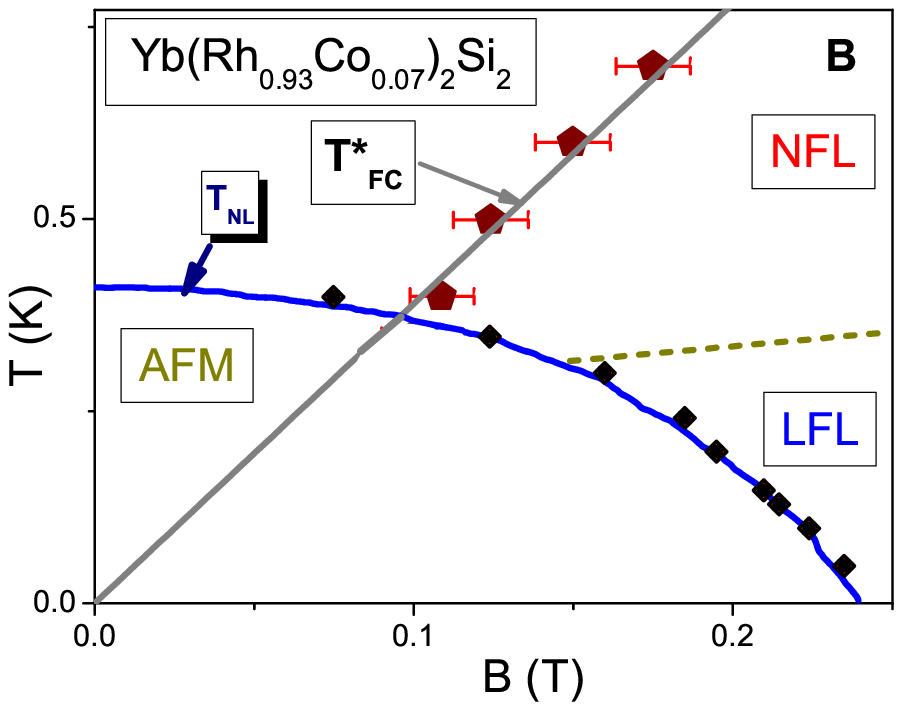}
\includegraphics [width=0.42\textwidth]{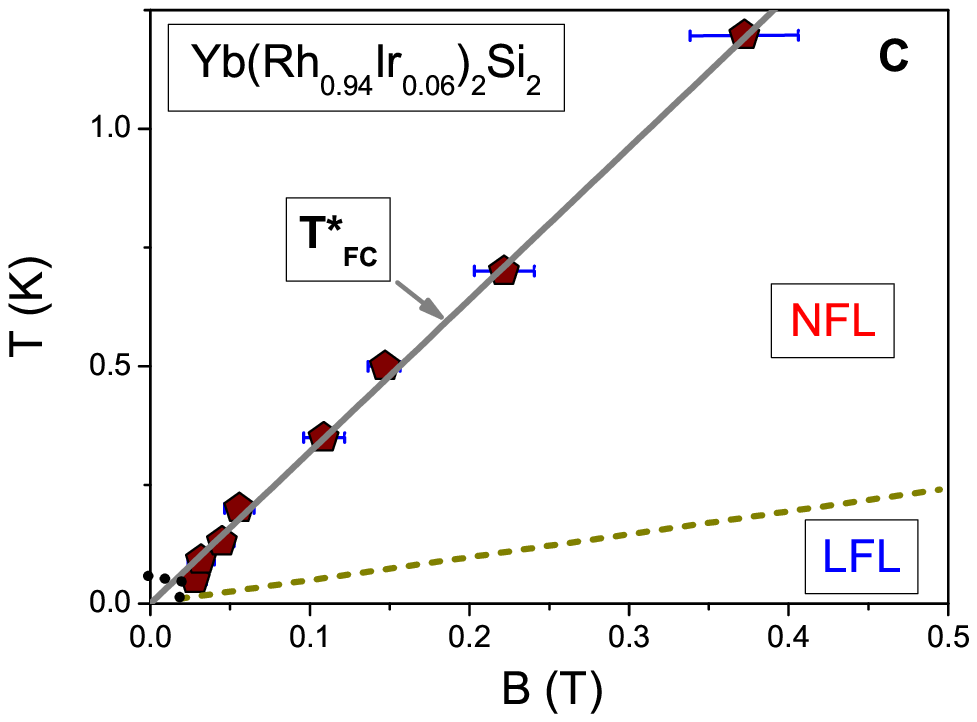}
\end{center}
\caption{(Color online). $T-B$ phase diagrams for the three HF
metals: $\rm YbRh_2Si_2$ (A); $\rm Yb(Rh_{0.93}Co_{0.07})_2Si_2$
(B); $\rm Yb(Rh_{0.94}Ir_{0.06})_2Si_2$ (C). In panels A and B, the
AF phase boundaries \cite{pss} are shown by the solid lines. In
panel B, the diamonds depict the measurements \cite{pss}. In panel
C, the phase boundary of possible phase transition is shown by
short dot curve. Pentagons correspond to the measurements of
$T^*(B)$ on the HF metals extracted from the analysis of
$\hat{\textbf{M}}$ as a function $B$ \cite{pss}. The solid straight
lines depict the transition temperature $T^*_{FC}(B)$. Dash lines
schematically denote the boundary of the NFL and LFL
regime.}\label{ff4}
\end{figure}
Panels (A,B,C) of Fig. \ref{ff4} are focused on the behavior of the
transition temperature $T^*(B)$ extracted from measurements of
kinks in $\hat{\textbf{M}}=\textbf{M}+(\textbf{dM}/dB)B$ as a
function of $B$ \cite{pss}, where $\textbf{M}$ is the
magnetization. Positions of the kinks is shown by pentagons in Fig.
\ref{ff4}. It is seen from Fig. \ref{ff4}, panel A, that at
$T\gtrsim T_{NL}(B=0)$, the transition temperature $T^*$ of $\rm
YbRh_2Si_2$ is well approximated by the line $T^*_{FC}$. As
mentioned above, upon using nonthermal tuning parameters like the
number density $x$, the NFL behavior can be destroyed and the LFL
one will be restored. In our simple model, the application of
positive pressure $P$ makes $x$ grow shifting the system from FCQPT
to the LFL state, as it is shown in Fig. \ref{fig03}, so that the
electronic system of $\rm YbRh_2Si_2$ moves into the shadow area
characterized by the LFL behavior at low temperatures. The new
location of the system, represented by $\rm
Yb(Rh_{0.93}Co_{0.07})_2Si_2$, is shown by the arrow pointing at
the solid square. We note that the positive chemical pressure in
the considered case is induced by $\rm Co$ substitution
\cite{stegjap,pss,stegnat,steg2}. As a result, the application of
magnetic field $B\simeq B_{c0}$ does not drive the system to its
FCQPT with the divergent effective mass because the QCP is already
destroyed by the positive pressure, as it is shown in Fig.
\ref{ff4}, panel B. Here $B_{c0}$ is the critical magnetic field
that eliminates the corresponding AF order. At $B>B_{c0}$ and
raising temperatures, the system, moving along the vertical arrow,
transits from the LFL regime to the NFL one. At relatively high
temperatures both $\rm YbRh_2Si_2$ and $\rm
Yb(Rh_{0.93}Co_{0.07})_2Si_2$ are in their paramagnetic state. As a
result, in that case $T^*$ is well approximated by the straight
line $T^*_{FC}$. This observation is in accordance with
experimental facts \cite{stegjap,pss} displayed in Fig. \ref{ff4},
panels A,B.

The system located above QCL exhibits the NFL behavior down to
lowest temperatures unless it is captured by a phase transition.
The behavior exhibiting by the system located above QCL is in
accordance with the experimental observations that study the
evolution of QCP in $\rm YbRh_2Si_2$ under the application of
negative chemical pressure induced with $\rm Ir$ substitution
\cite{stegjap,pss,stegnat,steg2}. We assume a simple model that the
application of negative pressure reduces $x$ and the electronic
system of $\rm YbRh_2Si_2$ moves from QCP to a new position over
QCL shown by the dash arrow in Fig. \ref{fig03}. Thus, the
electronic system of $\rm Yb(Rh_{0.94}Ir_{0.06})_2Si_2$ is located
at QCL and possesses a flat band, while the entropy includes $S_0$.
We predict that at lowering temperatures, the electronic system of
$\rm Yb(Rh_{0.94}Ir_{0.06})_2Si_2$ is captured by a phase
transition, since the NFL state above QCL is strongly degenerate
and the term $S_0$ should be eliminated. At diminishing
temperatures, this degeneracy is to be removed by some phase
transition which likely can be detected by the LFL state
accompanying it. The tentative boundary line of that transition is
shown by the short dot line in Fig. \ref{ff4}, panel C. It is also
seen from Fig. \ref{ff4}, panel C, that at elevated temperatures
$T^*$ is well approximated by the function $T^*_{FC}$.

Thus, at relatively high temperatures the transition temperature
$T^*_{FC}(B)$, shown in Fig. \ref{ff4}, panels A, B, and C by the
solid lines, coincides with $T^*(B)$ depicted by the pentagons.
For, as it was discussed above, the local properties of the systems
in question are given by their local free energy, formed by the NFL
region related with FC, as it is displayed in Fig. \ref{fig03}.

To conclude this section, we have for the first time theoretically
carried out a systematic study of the phase diagrams of strongly
correlated Fermi systems, including HF metals like $\rm
YbRh_2Si_2$, and considered the evolution of these diagrams in case
of the application of negative/positive pressure. We have observed
that at sufficiently high temperatures outside the AF phase the
transition temperature $T^*(B)$ follows an almost linear
B-dependence, and coincides with $T^*_{FC}(B)$, induced by the
presence of FC. The obtained results are in good agreement with
experimental facts \cite{stegjap,pss,stegnat,steg2}.

As we shall show in Sections 4 and 5, the phase diagram \ref{fig1}
of HF liquid describes the corresponding phase diagrams of quantum
spin liquids and quasicrystals as well, while the observed scaling
behavior is described by Eq. \eqref{UN2}. Thus, all these strongly
correlated compounds exhibit the uniform behavior, and allow us to
view that behavior as representing the main characteristic of the
new state of matter.

\section{Quantum spin liquid}

\begin{figure} [! ht]
\begin{center}
\vspace*{-0.2cm}
\includegraphics [width=0.47\textwidth]{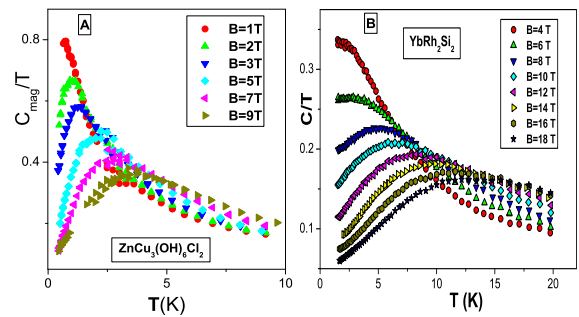}
\end{center}
\vspace*{-0.3cm} \caption{(Color online). Left panel: The specific
heat $C_{\rm mag}/T$ of QSL is extracted from measurements of the
specific heat on $\rm ZnCu_3(OH)_6Cl_2$ at different magnetic
fields shown in the legend \cite{19,20}. Right panel reports the
$T$-dependence of the electronic specific heat $C/T$ of $\rm
YbRh_2Si_2$ at different magnetic fields \cite{20} as shown in the
legend.}\label{FIG4}
\end{figure}

The first experimental observation of quantum spin liquid (QSL)
supporting exotic spin excitations - spinons - and carrying
fractional quantum numbers in the herbertsmithite $\rm
ZnCu_3(OH)_6Cl_2$ is reported in Ref. \cite{15}. QSL can be viewed
as an exotic quantum state composed of hypothetic particles such as
fermionic spinons which carry spin $1/2$ and no charge. The
herbertsmithite $\rm ZnCu_3(OH)_6Cl_2$ has been exposed as a
$S=1/2$ Heisenberg antiferromagnet on a perfect kagome lattice, see
Ref. \cite{16} for a recent review. In $\rm ZnCu_3(OH)_6Cl_2$, the
$\rm Cu^{2+}$ ions with $S=1/2$ form the triangular kagome lattice,
and are separated by nonmagnetic intermediate layers of $\rm Zn$
and $\rm Cl$ atoms. The planes of the $\rm Cu^{2+}$ ions can be
considered as two-dimensional (2D) layers with negligible magnetic
interactions along the third dimension. A simple kagome lattice has
a dispersionless topologically protected branch of the spectrum
with zero excitation energy that is the flat band \cite{17,18}. In
that case FCQPT can form a strongly correlated quantum spin liquid
(SCQSL) composed of fermions with zero charge, $S=1/2$, and the
effective mass $M^*_{\rm mag}$, occupying the corresponding Fermi
sphere with the Fermi momentum $p_F$. A comparison of the QSL
specific heat $C_{\rm mag}/T\propto M^*_{\rm mag}$ extracted from
measurements on $\rm ZnCu_3(OH)_6Cl_2$ with $C/T$ of $\rm
YbRh_2Si_2$ \cite{19,20,21} is shown in Fig. \ref{FIG4}. The
striking feature of the specific heat behavior is the strong
dependence on the magnetic field seen from the Figure. It is seen
that both $C_{\rm mag}/T$ and $C/T$ exhibit the same qualitative
behavior that allows us to view the herbertsmithite as insulator
with QSL, and QSL itself as SCQSL \cite{21,22,23}.

The schematic $T-B$ phase diagram of $\rm ZnCu_3(OH)_6Cl_2$ is
reported in Fig. \ref{fig1}. At $T=0$ and $B=0$ the system is near
FCQPT without tuning. It can also be shifted from FCQPT by the
application of magnetic field $B$. Magnetic field $B$ and
temperature $T$ play the role of the control parameters, driving it
from the NFL to LFL regions as shown by the vertical and horizontal
arrows. At fixed $B$ and increasing $T$ the system transits along
the vertical arrow from the LFL region to NFL one crossing the
transition region. On the contrary, at fixed $T$ increasing $B$
drives the system along the horizontal arrow from the NFL region to
LFL one. The inset to the phase diagram shown in Fig. \ref{fig04}
demonstrates the universal behavior of the normalized effective
mass $M^*_N$ versus normalized temperature $T_N$ as given by Eq.
\eqref{UN2}. It follows from Eq. \eqref{UN2}, and is seen from Fig.
\ref{fig1}, that the both width $W$ of the NFL and the width $T^W$
of the transition region, shown by the arrows in Fig. \ref{fig1},
tend to zero at diminishing $T$ and $B$ since $W\propto T^W \propto
T\propto B$.

\begin{figure} [! ht]
\begin{center}
\vspace*{-0.2cm}
\includegraphics [width=0.47\textwidth]{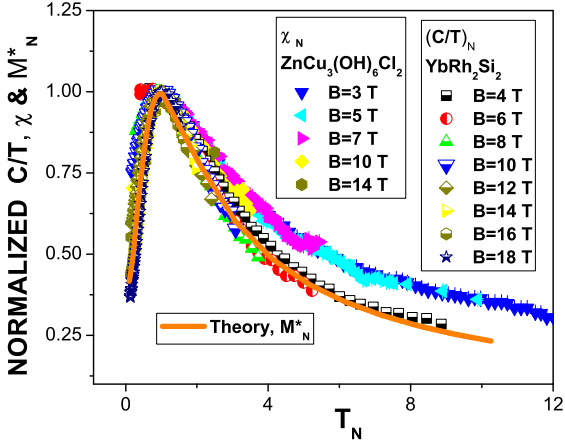}
\end{center}
\vspace*{-0.3cm} \caption{(Color online). The normalized effective
mass $M^*_N$ extracted from measurements of susceptibility $\chi$
\cite{24} on $\rm ZnCu_3(OH)_6Cl_2$ and $(C/T)$ on $\rm YbRh_2Si_2$
\cite{20}. Our calculations of $M^*_N$ are shown by solid
curve.}\label{FIG5}
\end{figure}
From Fig. \ref{FIG5} it is clearly seen that the data collected on
both $\rm ZnCu_3(OH)_6Cl_2$ \cite{24} and $\rm YbRh_2Si_2$
\cite{20} merge into the same curve, obeying the scaling behavior.
This demonstrates that the spin liquid of the herbertsmithite is
close to FCQPT and behaves like HF liquid of $\rm YbRh_2Si_2$ in
magnetic fields.

\begin{figure} [! ht]
\begin{center}
\includegraphics [width=0.47\textwidth]{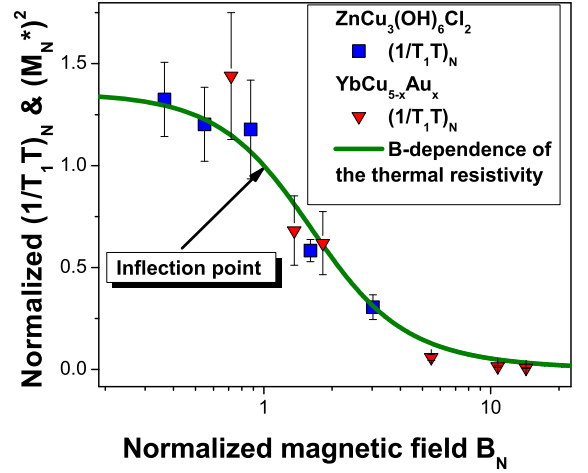}
\end{center}
\caption{(Color online). The relaxation properties of the
herbertsmithite versus those of HF metals. The normalized
spin-lattice relaxation rate $(1/T_1T)_N$ at fixed temperature as a
function of magnetic field: Solid squares correspond to data on
$(1/T_1T)_N$ extracted from measurements on $\rm ZnCu_3(OH)_6Cl_2$
\cite{imai}, while the solid triangles correspond to those
extracted from measurements on $\rm YbCu_{5-x}Au_{x}$ with $x=0.4$
\cite{carr}. The inflection point, representing the transition
region, where the normalization is taken is shown by the arrow. Our
calculations based on Eqs. \eqref{LF1} and Eq. \eqref{UN2} are
depicted by the solid curve, tracing the scaling behavior of
$(M^*_N)^2$ and representing the $B$-dependence of the thermal
resistivity $w$, see main text and Eq. \eqref{kap}.}\label{T1}
\end{figure}

Figure \ref{T1} displays the normalized spin-lattice relaxation
rates $(1/T_1T)_N$ at fixed temperature versus normalized magnetic
field $B_N$. It is seen from Fig. \ref{T1}, that the magnetic field
$B$ progressively reduces $1/T_1T$, and the spin-lattice relaxation
rate as a function of $B$ possesses an inflection point at some
$B=B_{inf}$ shown by the arrow. To clarify the universal scaling
behavior of QSL in the herbertsmithite  and in HF metals, we
normalize both the function $1/T_1T$ and the magnetic field.
Namely, we normalize $(1/T_1T)$ by its value at the inflection
point, and magnetic field is normalized by $B_{inf}$,
$B_N=B/B_{inf}$. Since $(1/T_1T)_N=(M^*_N)^2$ \cite{4,t1rel}, we
expect that different strongly correlated Fermi systems located
near FCQPT exhibit the same behavior of the normalized spin-lattice
relaxation rate. It is seen from Fig. \ref{T1}, that both the
herbertsmithite $\rm ZnCu_3(OH)_6Cl_2$ \cite{imai} and HF metal
$\rm YbCu_{5-x}Au_{x}$ \cite{carr} demonstrate similar behavior of
the normalized spin-lattice relaxation rate. As seen from Fig.
\ref{T1}, at $B\leq B_{inf}$ (or $B_N\leq1$), when the system in
its NFL region, the normalized relaxation rate $(1/T_1T)_N$ depends
weakly on the magnetic field, while at higher fields, as soon as
the system enters the LFL region, $(1/T_1T)_N$ diminishes in
agreement with both Eq. \eqref{UN2} and phase diagram \ref{fig1},
\begin{equation}\label{43}
(1/T_1T)_N=(M^*_N)^2\propto B^{-4/3}
\end{equation}
Thus, in accordance with the phase diagram shown in Fig. \ref{fig1}
and Eq. \eqref{43}, we conclude that the application of magnetic
field $B$ leads to crossover from the NFL to LFL behavior and to
the significant reduction in the relaxation rate.

As it was mentioned above, QSL plays a role of HF liquid framed
into the insulating compound. Thus, we expect that QSL in the
herbertsmithite behaves like the electronic liquid in HF metals if
the charge of an electron were zero. In that case, the thermal
resistivity $w$ of QSL is customarily  related to the thermal
conductivity $\kappa$
\begin{equation}\label{kap}
w=\frac{L_0T}{\kappa}=w_0+A_wT^2.
\end{equation}
The resistivity $w$ behaves like the electrical resistivity
$\rho=\rho_0+A_{\rho}T^2$ of the electronic liquid, since $A_w$
represents the contribution of spinon-spinon scattering to the
thermal transport, being analogous to the contribution $A_{\rho}$
to the charge transport by electron-electron scattering. Here,
$L_0$ is the Lorenz number, $\rho_0$ and $w_0$ are residual
resistivity of electronic liquid and QSL, respectively, and the
coefficients $A_w\propto (M^*_{\rm mag})^2$ and $A_{\rho}\propto
(M^*)^2$ \cite{4}. Thus, in the LFL region the coefficient $A_w$ of
the thermal resistivity of QSL under the application of magnetic
fields at fixed temperature behaves like the spin-lattice
relaxation rate shown in Fig. \ref{T1}, $A_w(B)\propto
1/T_1T(B)\propto (M^*(B)_{\rm mag})^2$, while in the LFL region at
fixed magnetic fields the thermal conductivity $\kappa$ is a linear
function of temperature, $\kappa\propto T$.

Study of the thermal resistivity $w$ given by Eq. \eqref{kap}
allows one to reveal spinons as itinerant excitations. It is
important that $w$ is not contaminated by contributions coming from
localized excitations by impurity effects. The temperature
dependence of thermal resistivity $w$ represented by the finite
term $w_0$ directly shows that the behavior of QSL is similar to
that of metals, and there is a finite residual term $\kappa/T$ in
the zero-temperature limit of $\kappa$. The presence of this term
immediately proves that there are gapless excitation associated
with the property of normal metals, in which gapless electrons
govern the heat transport. The finite $w_0$ means that in QSL both
$k/T$ and $C_{mag}/T\propto M^*_{\rm mag}$ remain nonzero at
$T\to0$. Therefore, gapless spinons, forming the Fermi surface,
govern the specific heat and the transport. Key information on the
nature of spinons is further provided by the $B$-dependence of the
coefficient $A_w$. The specific $B$-dependence of the resistivity
$w(B)\propto(M^*_{\rm mag})^2$, shown in Fig. \ref{T1} and given by
Eq. \eqref{43}, would establish the behavior of QSL as SCQSL. We
note that the heat transport is polluted by the phonon
contribution. On the other hand, the phonon contribution is hardly
influenced by the magnetic field $B$. Therefore, we expect the
$B$-dependence of the heat conductivity to be governed by
$A_w(B,T)$. Consider the approximate relation,
\begin{eqnarray}
\nonumber 1&-&\frac{A_w(B,T)}{A_w(0,T)}=
1-\left(\frac{M^*(B,T)_{\rm mag}}{M^*(0,T)_{\rm mag}}\right)^2\\
&\simeq&a(T)\frac{\kappa(B,T)-\kappa(0,T)}{\kappa(0,T)}\equiv
a(T)I(B,T),\label{TR}
\end{eqnarray}
where the coefficient $a(T)$ is $B$-independent. To derive
\eqref{TR}, we employ Eq. \eqref{kap}, and obtain
\begin{equation}\label{TR1}
\frac{\kappa}{L_0T}=\frac{1}{w_0+A_wT^2}+bT^2.
\end{equation}
Here, the term $bT^2$ describes the phonon contribution to the heat
transport. Upon carrying out simple algebra and assuming that
$[1-A_w(B,T)/A_w(0,T)]\ll 1$, we arrive at Eq. \eqref{TR}. It is
seen from Fig. \ref{T1}, that the effective mass $M^*_N(B)\propto
M^*_{\rm mag}(B)$ is a diminishing function of magnetic field $B$.
Then, it follows from Eqs. \eqref{43} and \eqref{TR} that the
function $I(B,T)=[\kappa(B,T)-\kappa(0,T)]/\kappa(0,T)$ increases
at elevated field $B$.
\begin{figure} [! ht]
\begin{center}
\vspace*{-0.2cm}
\includegraphics [width=0.47\textwidth]{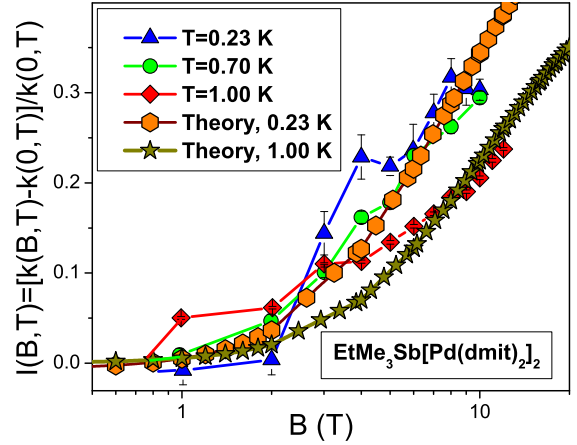}
\end{center}
\vspace*{-0.3cm} \caption{(Color online). Magnetic field $B$
dependence of the thermal conductivity $I(B,T)$ measured on the
organic insulator $\rm EtMe{_3}Sb[Pd(dmit)_{2}]_{2}$ and
standardized by the zero field value $\kappa$,
$I(B,T)=[\kappa(B,T)-\kappa(B=0,T)]/\kappa(B=0,T)$ at temperatures
shown in the legend \cite{scqsl,chqs1}. Our calculations are based
on Eq. \eqref{TR} and shown by pentagons and stars.}\label{kappa}
\end{figure}

\begin{figure} [! ht]
\begin{center}
\vspace*{-0.2cm}
\includegraphics [width=0.47\textwidth]{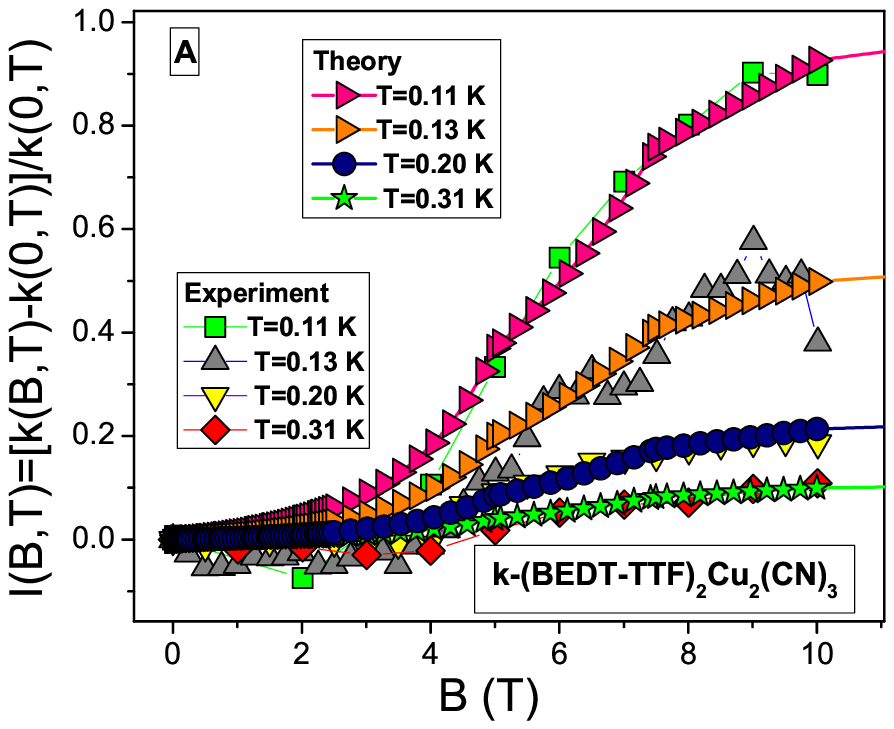}
\includegraphics [width=0.47\textwidth]{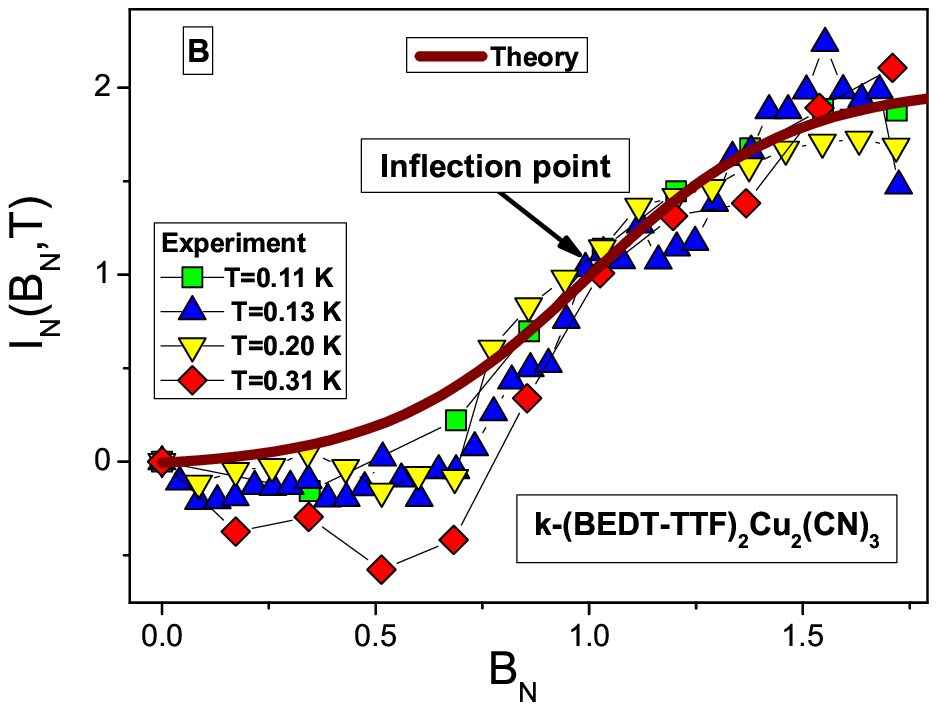}
\end{center}
\vspace*{-0.3cm} \caption{(Color online). Panel {\bf A}: Magnetic
field $B$ dependence of the thermal conductivity $I(B,T)$ measured
on the organic insulator $\rm\kappa-(BEDT-TTF)_2Cu_2(CN)_3$ and
standardized by the zero field value $\kappa$,
$I(B,T)=[\kappa(B,T)-\kappa(B=0,T)]/\kappa(B=0,T)$ at temperatures
shown in the legend \cite{chqs1}. Our calculations are based on Eq.
\eqref{TR} and shown by geometrical figures as it is displayed in
the legend. Panel {\bf B}: Normalized thermal conductivity
$I_N(B_N,T)$ as function of the normalized magnetic field $B_N$.
The inflection point is shown by the arrow. Our calculations based
on Eq. \eqref{TR} are depicted by the solid curve.}\label{kappa1}
\end{figure}

Recent measurements of $\kappa(B)$ on the organic insulators $\rm
EtMe{_3}Sb[Pd(dmit)_{2}]_{2}$ and
$\rm\kappa-(BEDT-TTF)_2Cu_2(CN)_3$ \cite{scqsl,chqs1} are displayed
in Figs. \ref{kappa} and \ref{kappa1}. The measurements show that
the heat is carried by phonons and QSL, for the heat conductivity
is well fitted by $\kappa/T=a_1+a_2T^2$, where $a_1$ and $a_2$ are
constants. The finite $a_1$ term implies that spinon excitations
are gapless in $\rm EtMe{_3}Sb[Pd(dmit)_{2}]_{2}$, while in
$\rm\kappa-BEDT-TTF)_2Cu_2(CN)_3$ gapless excitations are under
debate \cite{chqs1}. A simple estimation indicates that the
ballistic propagation of spinons seems to be realized in the case
of $\rm EtMe{_3}Sb[Pd(dmit)_{2}]_{2}$ \cite{scqsl,chqs1}. It is
seen from Figs. \ref{kappa} and \ref{kappa1}, panel {\bf A}, that
the normalized data
$I(B,T)=[\kappa(B,T)-\kappa(B=0,T)]/\kappa(B=0,T)$ demonstrate a
strong $B$-dependence, namely the field dependence shows an
increase of thermal conductivity for rising fields $B$. Such a
behavior is in agreement with Eq. \eqref{43} and Fig. \ref{T1}
which demonstrate that $(M^*(B)_{mag})^2$ is diminishing function
of $B$. As a result, it follows from Eq. \eqref{TR} that $I(B,T)$
is an increasing function of $B$. Our calculations based on Eqs.
\eqref{TR} and \eqref{LF1} are depicted by geometrical figures in
Figs. \ref{kappa} and \ref{kappa1}, panel {\bf A}. Since we cannot
calculate the parameter $a(T)$ entering Eq. \eqref{TR} we use it as
a fitting parameter. Temperature $T$ was also used to fit the data
at temperatures shown in the legend in Figs. \ref{kappa} and
\ref{kappa1}. It is seen from Fig. \ref{kappa1}, panel {\bf A},
that $I(B,T)$ as a function of $B$ possesses an inflection point at
some $B=B_{inf}$. To reveal the scaling behavior of QSL in
$\rm\kappa-(BEDT-TTF)_2Cu_2(CN)_3$, we normalize both the function
$I(B,T)$ and the magnetic field by their values at the inflection
points, as it was done in the case of $(1/T_1T)$, see Fig.
\ref{T1}. In that case we get rid of the factor $a(T)$, entering
Eq. \eqref{TR}, and our calculations do not have any fitting
parameters. It is seen from Fig. \ref{kappa1}, panel {\bf B}, that
the normalized $I_N(B_N,T)$ exhibits the scaling behavior and
becomes a function of a single variable $B_N$. Our calculations
show that $I_N(B_N,T)$ extracted from measurements on $\rm
EtMe{_3}Sb[Pd(dmit)_{2}]_{2}$ exhibits the scaling behavior as
well. It is seen from both Figs. \ref{kappa} and \ref{kappa1}, that
our calculations are in good overall agreement with the
experimental facts and there is no need to suppose the existence of
additional magnetic excitations activated by the application of
magnetic field in order to explain the growth of $I(B,T)$ at
elevated $B$ \cite{scqsl,chqs1}.

Important signature of electron liquid in HF metals is excitations
--- quasiparticles carrying electron quantum numbers and
characterized by the effective mass $M^*_{\rm mag}$. Neutron
scattering measurements of the dynamic spin susceptibility
$\chi({\bf q},\omega,T)=\chi{'}({\bf q},\omega,T)+i\chi{''}({\bf
q},\omega,T)$ as a function of momentum $q$, frequency $\omega$ and
and temperature $T$ play important role when identifying the
properties of quasiparticles.
\begin{figure} [! ht]
\begin{center}
\vspace*{-0.2cm}
\includegraphics [width=0.47\textwidth]{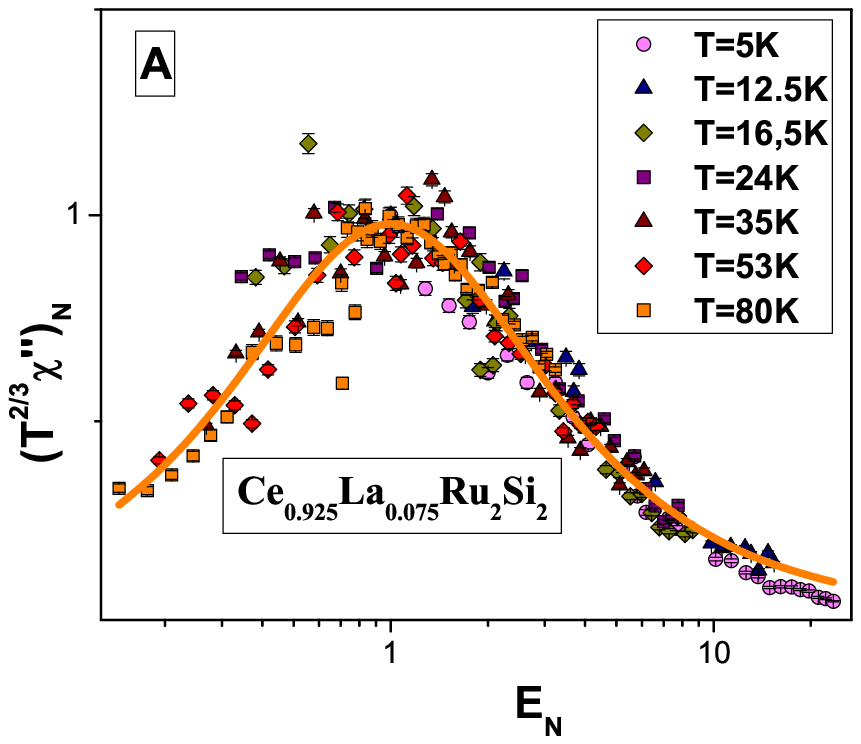}
\includegraphics [width=0.47\textwidth]{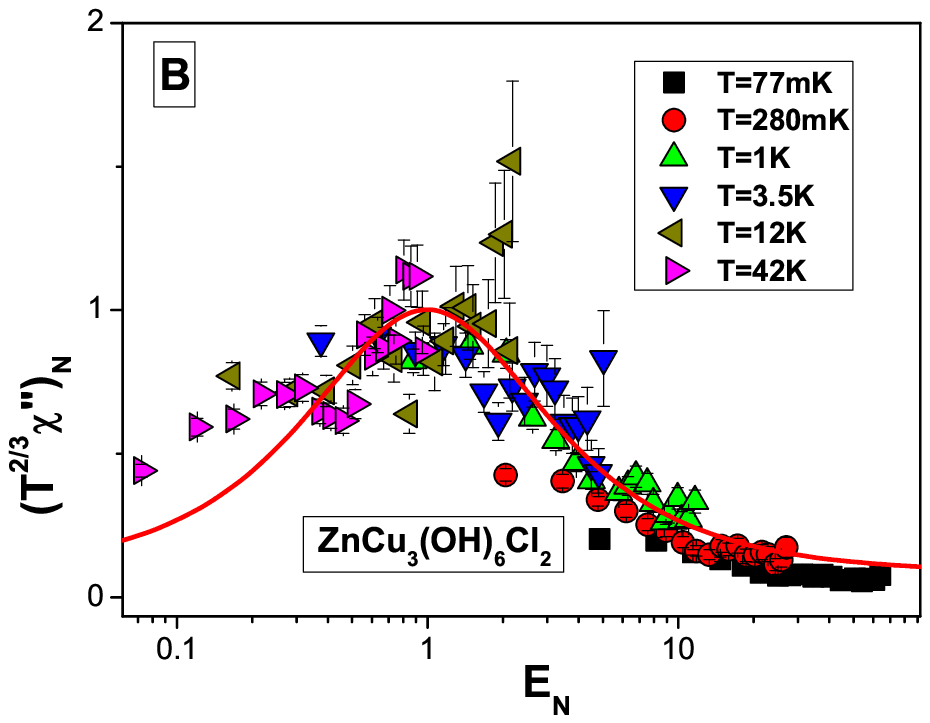}
\end{center}
\vspace*{-0.3cm} \caption{(Color online). Panel {\bf A}: Normalized
$(T^{2/3}\chi'')_N$ plotted against the dimensionless variable
$E_N$. The data are extracted from measurements on $\rm
Ce_{0.925}La_{0.075}Ru_2Si_2$ \cite{25}. Panel {\bf B}:
$(T^{2/3}\chi'')_N$ versus $E_N$. The data are extracted from
measurements on $\rm ZnCu_3(OH)_6Cl_2$ \cite{24}. Our calculations
are shown by the solid curves.}\label{FIG6}
\end{figure}
At low temperatures, such measurements allow one to reveal that
quasiparticles excitations form a continuum, and populate an
approximately flat band crossing the Fermi level. In that case the
normalized susceptibility $(T^{2/3}\chi'')_{N}$ exhibits a scaling
behavior as a function of the dimensionless variable $E_N$
\cite{23}. The panel {\bf A} of Fig. \ref{FIG6} reports
$(T^{2/3}\chi'')_{N}$ extracted from measurements of the inelastic
neutron scattering spectrum on the HF metal $\rm
Ce_{0.925}La_{0.075}Ru_2Si_2$ \cite{25}. The data
$(T^{2/3}\chi'')_{N}$ exhibit the scaling behavior over three
orders of the variation of both the function and the variable. The
scaled data obtained in measurements on a such quite different
strongly correlated system as $\rm ZnCu_3(OH)_6Cl_2$ \cite{24} are
shown in the panel {\bf B}. It is seen that our calculations shown
by the solid curves are in good agreement with the experimental
facts collected both on $\rm Ce_{0.925}La_{0.075}Ru_2Si_2$ and $\rm
ZnCu_3(OH)_6Cl_2$ over almost three orders of the scaled variables
\cite{jetp}. Thus, the spin excitations in $\rm ZnCu_3(OH)_6Cl_2$
exhibit the same behavior as electron excitations of the HF metal
$\rm Ce_{0.925}La_{0.075}Ru_2Si_2$, and, therefore form a
continuum. This observation of the continuum is of great importance
since it clearly reveals the presence of SCQSL in the
herbertsmithite, as it was later confirmed by direct experimental
observation \cite{15}.

In this section, we have considered the non-Fermi liquid behavior
and the scaling one of such strongly correlated Fermi systems as
insulators $\rm ZnCu_3(OH)_6Cl_2$, $\rm
EtMe{_3}Sb[Pd(dmit)_{2}]_{2}$, $\rm\kappa-(BEDT-TTF)_2Cu_2(CN)_3$,
and HF metals $\rm Ce_{0.925}La_{0.075}Ru_2Si_2$, $\rm
YbCu_{5-x}Au_{x}$, and $\rm YbRh_2Si_2$. We have shown that these
are described within the frame of the theory of FC. Our
calculations are in a good agreement with the experimental data,
and allow us to identify the low-temperature behavior of $\rm
ZnCu_3(OH)_6Cl_2$, $\rm\kappa-(BEDT-TTF)_2Cu_2(CN)_3$, and $\rm
EtMe{_3}Sb[Pd(dmit)_{2}]_{2}$ as determined by SCQSL.  The same
behavior is observed in the heavy fermion metals. Thus, $\rm
ZnCu_3(OH)_6Cl_2$, $\rm\kappa-BEDT-TTF)_2Cu_2(CN)_3$, and $\rm
EtMe{_3}Sb[Pd(dmit)_{2}]_{2}$ can be viewed as a new type of
strongly correlated electrical insulator that possesses properties
of heavy-fermion metals with one exception: it resists the flow of
electric charge, supporting the spin current formed by spinons.

\section{Quasicrystals}

New materials named quasicrystals (QCs) and characterized by
noncrystallographic rotational symmetry and quasiperiodic
translational properties have attracted scrutiny \cite{26}. Study
of quasicrystals may shed light on the most basic notions related
to the quantum critical state observed in HF metals. Experimental
measurements on the gold-aluminium-ytterbium quasicrystal $\rm
Au_{51}Al_{34}Yb_{15}$ have revealed a quantum critical behavior
with the unusual exponent $\beta\simeq0.51$ defining the divergency
of the magnetic susceptibility $\chi\propto T^{-\beta}$ at $T\to0$
\cite{27}. The measurements have also exposed that the observed NFL
behavior transforms into the LFL one under the application of a
tiny magnetic field $B$. All these facts challenge theory to
explain a quantum criticality of the gold-aluminum-ytterbium QC. In
case of QCs electrons occupy a new class of states denoted as
"critical states", neither being extended nor localized. Associated
with these critical states, characterized by an extremely
degenerate confined wave function, are the so-called "spiky" DOS
\cite{28}. These predicted DOS are corroborated by experiments
revealing that single spectra of the local DOS demonstrate the
spiky DOS \cite{29}, which form flat bands \cite{30}. As a result,
we assume that the electronic system of some quasicrystals is
located at FCQPT without tuning.

\begin{figure} [! ht]
\begin{center}
\vspace*{-0.2cm}
\includegraphics [width=0.47\textwidth]{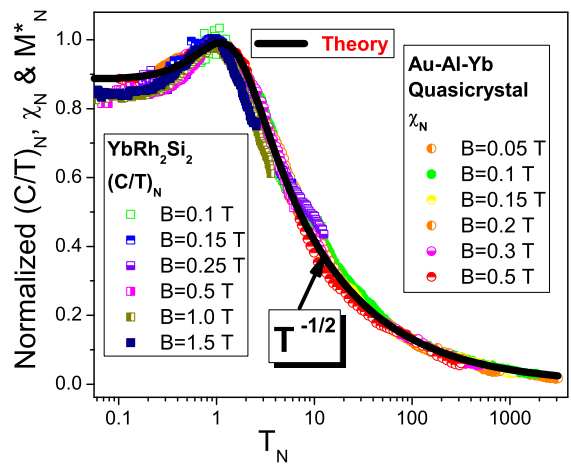}
\end{center}
\vspace*{-0.3cm} \caption{(Color online). The normalized specific
heat $(C/T)_N$ and the normalized magnetic susceptibility $\chi_N$
extracted from measurements in magnetic fields $B$ on $\rm
YbRh_2Si_2$ \cite{pikul} and on $\rm Au_{51}Al_{34}Yb_{15}$
\cite{27}, respectively. The corresponding magnetic fields are
listed in the legends. The arrow shows the $T^{-1/2}$ regime taking
place at $T_N>1$. Our calculations are depicted by the solid curve
tracing the scaling behavior of $(C/T)_N=\chi_N=M^*_N$ given by Eq.
\eqref{UN2}.}\label{FIG7_1}
\end{figure}

We now investigate the behavior of $\chi$ as a function of
temperature at fixed magnetic fields. The effective mass $M^*(T,B)$
can be measured in experiments, for $M^*(T,B)\propto \chi$, where
$\chi$ is the ac or dc magnetic susceptibility. If the
corresponding measurements are carried out at fixed magnetic field
$B$ then, as it follows from Eq. \eqref{UN2}, $\chi$ reaches the
maximum $\chi_{M}$ at some temperature $T_{M}$. Upon normalizing
both $\chi$ and the specific heat $C/T$ by their peak values at
each field $B$  and the corresponding temperatures by $T_{M}$, we
observe from Eq. \eqref{UN2} that all the curves are to merge into
a single one, thus demonstrating a scaling behavior typical for HF
metals \cite{4}. As seen from Fig. \ref{FIG7_1}, $\chi_N$ extracted
from measurements on $\rm Au_{51}Al_{34}Yb_{15}$ \cite{27} shows
the scaling behavior given by Eq. \eqref{UN2} and agrees well with
our calculations shown by the solid curve over four orders of
magnitude in the normalized temperature. It is also seen, that
$\chi_N$ agrees well with the normalized $(C/T)_N$ extracted from
measurements in magnetic fields on $\rm YbRh_2Si_2$ \cite{pikul}.
At $T_N>1$ the susceptibility $\chi_N$ exhibits the $T^{-\beta}$
regime, $\chi_N(T)\propto T^{-\beta}$, with $\beta=1/2$.

The schematic phase diagram of the gold-aluminum-ytterbium QC $\rm
Au_{51}Al_{34}Yb_{15}$ is reported in Fig. \ref{fig1}. The magnetic
field $B$ plays the role of the control parameter, driving the
system outwards FCQPT that occurs at $B=0$ and $T=0$ without
tuning, since the QC critical state is formed by singular density
of states \cite{27,28,29,30}. It follows from Eq. \eqref{UN2} and
is seen from Fig. \ref{fig1}, that at fixed temperatures the
increase of $B$ drives the system along the horizontal arrow from
NFL state to LFL one. On the contrary, at fixed magnetic field and
increasing temperatures the system transits along the vertical
arrow from LFL state to NFL one. The $T^{-\beta}$ regime with
$\beta=1/2$ is marked as NFL since contrary to the LFL case, where
the effective mass is constant, the effective mass depends strongly
on temperature. Thus, the temperatures $T^*_{FC}\simeq T_{M}$,
shown by the arrow in Fig. \ref{fig1}, can be regarded as the
transition regime between LFL and NFL states. It is seen, that the
common width of the LFL region $W$ and the transition one $T^W$ is
proportional to $T$. These theoretical results are in good
agreement with the experimental facts \cite{27}.

\begin{figure} [! ht]
\begin{center}
\vspace*{-0.2cm}
\includegraphics [width=0.47\textwidth]{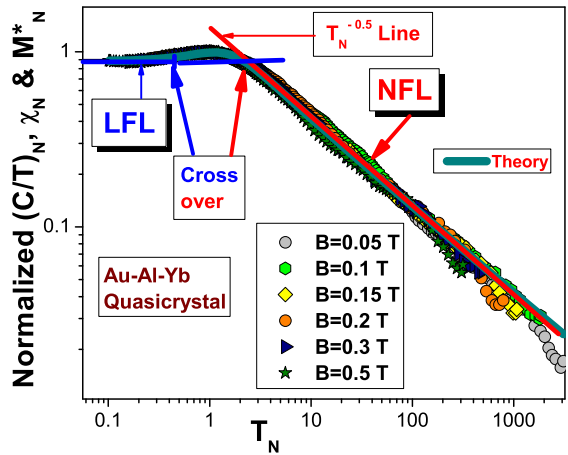}
\end{center}
\vspace*{-0.3cm} \caption{(Color online). Temperature dependence on
the double logarithmic scale of the normalized effective mass
$M^*_N$ versus normalized temperature $T_N$ extracted from
measurements of the magnetic susceptibility $\chi$ on the
quasicrystal $\rm Au_{51}Al_{35}Yb_{14}$ at different magnetic
fields \cite{27} shown in the legend. The LFL region and NFL one
are shown by the solid and dashed arrows, respectively. The
crossover region is depickted by two arrows. The solid line marked
by the NFL displays $M^*_N\propto T_N^{-0.5}$
behavior.}\label{FIG7}
\end{figure}

In order to validate the phase diagram displayed in Fig.
\ref{fig1}, we focus on the LFL, NFL and the transition LFL-NFL
regions exhibited by the QC. To this end, we display in Fig.
\ref{FIG7} the normalized $\chi_N$ on the double logarithm scale.
As seen from Fig. \ref{FIG7}, $\chi_N$ extracted from the
measurements is not a constant, as would be for a LFL. The two
regions (the LFL region and NFL one), separated by the transition
region, as depicted by the hatched area in Fig. \ref{fig1}, are
clearly seen in Fig. \ref{FIG7} illuminating good agreement between
the theory and measurements. The straight lines in Fig. \ref{FIG7}
outline both the LFL and NFL behaviors of $\chi_N\propto const$ and
$\chi_N\propto T_N^{-1/2}$. These straight lines in Fig. \ref{FIG7}
are in good agreement with the behavior of $M^*_N$ given by Eq.
\eqref{UN2} with $n=5/2$. It is also seen from Fig. \ref{FIG7} that
a tiny magnetic field of $B=0.05$ T destroys the NFL behavior
hereby driving the system to the LFL region \cite{14}. It is seen
that our calculations shown by the solid curve are in good
agreement with the experimental facts over four orders of magnitude
in the normalized temperature.

As a result, we conclude that $\rm Au_{51}Al_{34}Yb_{15}$
quasicrystal exhibits the typical scaling behavior of its
thermodynamic properties, and belongs to the famous family of HF
metals, while the quantum critical physics of the quasicrystal is
universal, and emerges regardless of the underlying microscopic
details of the quasicrystal.

\section{Summary}

The condensate state in Bose interacting systems, introduced by S.
T. Belyaev in his famous papers, forms the system's properties. It
turns out that the fermion condensate takes place in many
compounds, and generates the non-Fermi liquid behavior by forming
flat bands. We have for the first time theoretically carried out a
systematic study of the phase diagrams of strongly correlated Fermi
systems, including HF metals, insulators with strongly correlated
quantum spin liquid, and quasicrystals, and have demonstrated that
these diagram have universal features. The obtained results are in
good agreement with experimental facts. We have shown both
analytically and using arguments based entirely on the experimental
grounds that the data collected on very different heavy-fermion
compounds, such as HF metals, compounds with quantum spin liquid
and quasicrystals, have a universal scaling behavior, and materials
with strongly correlated fermions can unexpectedly have a uniform
behavior in spite of their microscopic diversity. Thus, the quantum
critical physics of different heavy-fermion compounds is universal,
and emerges regardless of the underlying microscopic details of the
compounds. This uniform behavior, induced by the universal quantum
critical physics, allows us to view it as the main characteristic
of the new state of matter. Our analysis of strongly correlated
systems is in the context of salient experimental results, and our
calculations of the non-Fermi liquid behavior are in good agreement
with a broad variety of experimental facts. Our theoretical
analysis of numerous experimental facts shows that the theory of
fermion condensation develops completely good description of the
NFL behavior of strongly correlated Fermi systems. Moreover, the
fermion condensate can be considered as the universal reason for
the NFL behavior observed in various HF metals, liquids, insulators
with quantum spin liquids, and quasicrystals.

\section{acknowledgement}

This paper is written as invited review dedicated to 90th
anniversary of S. T. Belyaev birthday.

KGP acknowledges funding from the Ural Branch of the Russian
Academy of Sciences, basic research program no. 12-U-1-1010, the
Presidium of the Russian Academy of Sciences, program 12-P1-1014,
and RTP UrB RAS project P3.


\begin{thebibliography}{99}

\bibitem{bel1} S. T. Belyaev, Sov. Phys. JETP {\bf 34}, 417 (1958).

\bibitem{bel2} S. T. Belyaev, Sov. Phys. JETP {\bf 34}, 433 (1958).

\bibitem{ks} V. A. Khodel and V. R. Shaginyan,
JETP Lett. {\bf 51}, 553 (1990).

\bibitem{pr1} V. A.
Khodel, V. R. Shaginyan, and V. V. Khodel, Phys. Rep. {\bf 249}, 1
(1994).

\bibitem{4} V. R. Shaginyan, M. Ya. Amusia, A. Z. Msezane, and K. G. Popov,
Phys. Rep. {\bf 492}, 31 (2010).

\bibitem{vol} G.  E. Volovik, JETP Lett. {\bf 53}, 222 (1991).

\bibitem{volovik2} G. E. Volovik, Springer Lecture Notes in Phys. {\bf
718}, 31 (2007).

\bibitem{hwk}  P. Hohenberg and W. Kohn, Phys. Rev. {\bf 136}, B864
(1965).

\bibitem{wks}  W. Kohn and L. J. Sham, Phys. Rev. {\bf 140}, A1133
(1965).

\bibitem{aplh} M. V. Zverev, V. A. Khodel, V. R. Shaginyan, and M. Baldo,
JETP Lett. {\bf 65}, 863 (1997).

\bibitem{1_1} L. D. Landau, Sov. Phys. JETP {\bf 5}, 101 (1957).

\bibitem{1_2} L. D. Landau, Sov. Phys. JETP {\bf 8}, 70 (1959).

\bibitem{lanl}  E. M. Lifshitz and L. P. Pitaevskii,
{\it Statistical Physics}, Pt. 2, Pergamon Press, Oxford, 1980.

\bibitem{UBH} V. R. Shaginyan, JETP Lett.  {\bf 79}, 286
(2004).

\bibitem{UFN} V. R. Shaginyan, M. Ya. Amusia, and K. G. Popov, Physics-Uspekhi
{\bf 50}, 563 (2007).

\bibitem{pom} I. Ya. Pomeranchuk, Sov. Phys. JETP {\bf 8}, 361 (1958).

\bibitem{noz} P. Nozi\`{e}res, J. Phys. I France {\bf 2}, 443 (1992).

\bibitem{8} A. Casey, H. Patel, J. Nyeki, B.P. Cowan, and J.
Saunders, J. Low Temp. Phys. {\bf 113}, 293 (1998).

\bibitem{oesb} N. Oeschler, S. Hartmann, A. P. Pikul, C. Krellner,
C. Geibel, and F. Steglich, Physica B {\bf 403}, 1254 (2008).

\bibitem{dft373} V. R. Shaginyan, M. Ya. Amusia, and K. G. Popov,
Phys. Lett. A {\bf 373}, 2281 (2009).

\bibitem{14} V. R. Shaginyan, A. Z. Msezane, K. G. Popov,
G. S. Japaridze, and V. A. Khodel,  Phys. Rev. B {\bf 87}, 245122
(2013).

\bibitem{asp} S.  A. Artamonov, Yu. G. Pogorelov,
and V. R. Shaginyan, JETP Lett. {\bf 68}, 942 (1998).

\bibitem{zvbld} M. V. Zverev and M. Baldo, JETP {\bf 87}, 1129 (1998).

\bibitem{khodb} V. A. Khodel, J. W.
Clark, and M. V. Zverev, Phys. Rev. B {\bf 78}, 075120 (2008).

\bibitem{yakov} V. A. Khodel, M. V. Zverev, and V. M. Yakovenko,
Phys. Rev. Lett. {\bf 95}, 236402 (2005).

\bibitem{shag} V. R. Shaginyan, Phys. Atom. Nucl.
{\bf 74}, 1107 (2011).

\bibitem{mig100}  V. A. Khodel, J. W. Clark, and M. V. Zverev,
Phys. Atom. Nucl. {\bf 74}, 1230 (2011).

\bibitem{takah} D. Takahashi,
S. Abe, H. Mizuno, D. A. Tayurskii, K. Matsumoto, H. Suzuki, and Y.
Onuki,  Phys. Rev. B {\bf 67}, 180407 (2003).

\bibitem{geg} P. Gegenwart, J. Custers, C. Geibel,
K. Neumaier, T. Tayama, K. Tenya, O. Trovarelli, and F. Steglich,
Phys. Rev. Lett. {\bf 89}, 056402 (2002).

\bibitem{jrho}  V. R. Shaginyan, A. Z. Msezane, K. G. Popov, J. W. Clark, M. V.
Zverev, and V. A. Khodel, JETP Lett. {\bf 96}, 397 (2012).

\bibitem{steg1} P. Gegenwart, T. Westerkamp, C. Krellner, Y. Tokiwa, S. Paschen,
C. Geibel, F. Steglich, E. Abrahams, and Q. Si, Science {\bf 315},
969 (2007).

\bibitem{jetp90} V. R. Shaginyan, M. Ya. Amusia, K. G. Popov, and S. A.
Artamonov,  JETP Lett. {\bf 90}, 47 (2009).

\bibitem{pss} M. Brando, L. Pedrero, T. Westerkamp, C. Krellner, P. Gegenwart,
C. Geibel, and F. Steglich, Phys. Status Solidi B {\bf 459}, 285
(2013).

\bibitem{stegjap} Y. Tokiwa, P. Gegenwart, C. Geibel, and F.
Steglich, J. Phys. Soc. Jpn. {\bf 78}, 123708 (2009).

\bibitem{stegnat} S. Friedemann, T. Westerkamp, M. Brando, N. \"Oeschler, S. Wirth,
P. Gegenwart, C. Krellner, C. Geibel, and F. Steglich, Nat. Phys.
{\bf 5}, 465 (2009).

\bibitem{steg2} J. Custers, P. Gegenwart, C. Geibel, F. Steglich, P. Coleman, and S.
Paschen, Phys. Rev. Lett. {\bf 104}, 186402 (2010).

\bibitem{15} T.-H. Han, J. S. Helton, S. Chu, D. G. Nocera,
J. A. Rodriguez-Rivera, C. Broholm, and Y. S. Lee, Nature {\bf
492}, 406 (2012).

\bibitem{16} F. Bert and P. Mendels, J. Phys. Soc. Jpn. {\bf 79}, 011001
(2010).

\bibitem{17} D. Green, L. Santos, and C. Chamon, Phys. Rev. B {\bf 82}, 075104
(2010).

\bibitem{18} T. T. Heikkil\"a, N. B. Kopnin, and G. E. Volovik,
JETP Lett. {\bf 94}, 233 (2011).

\bibitem{19} M. A. deVries, K. V. Kamenev, W. A. Kockelmann, J. Sanchez-Benitez,
and A. Harrison Phys. Rev. Lett. {\bf 100}, 157205 (2008).

\bibitem{20} P. Gegenwart, Y. Tokiwa, T. Westerkamp, F. Weickert,
J. Custers, J. Ferstl, C. Krellner, C. Geibel, P. Kerschl, K.-H.
M\"uller, and F. Steglich, New J. Phys. {\bf 8}, 171 (2006).

\bibitem{21} V. R. Shaginyan, A. Z. Msezane, K. G. Popov, G. S. Japaridze, and V. A.
Stephanovich, Europhys. Lett. {\bf 97}, 56001 (2012).

\bibitem{22} V. R. Shaginyan, A. Z. Msezane, and K. G. Popov,
Phys. Rev. B {\bf 84}, 060401(R) (2011).

\bibitem{23} V. R. Shaginyan, A. Z. Msezane, K. G. Popov, and V. A. Khodel,
Phys. Lett. A {\bf 376}, 2622 (2012).

\bibitem{24} J. S. Helton, K. Matan, M. P. Shores, E. A. Nytko,
B. M. Bartlett, Y. Qiu, D. G. Nocera, and Y. S. Lee, Phys. Rev.
Lett. {\bf 104}, 147201 (2010).

\bibitem{t1rel} V. R. Shaginyan, A. Z. Msezane, K. G. Popov, and V. A. Stephanovich,
Phys. Lett. A {\bf 373}, 3783 (2009).

\bibitem{imai} T. Imai, E. A. Nytko, B. M. Bartlett, M. P. Shores, and D. G. Nocera
Phys. Rev. Lett. {\bf 100}, 077203 (2008).

\bibitem{carr} P. Carretta, R. Pasero, M. Giovannini, and C. Baines,
Phys. Rev. B {\bf 79}, 020401(R) (2009).

\bibitem{scqsl} H. M. Yamamoto, N. Nakata, Y. Senshu, M. Nagata,
H. M. Yamamoto, R. Kato, T. Shibauchi, and Y. Matsuda, Science {\bf
328}, 1246 (2010).

\bibitem{chqs1} M. Yamashita, T. Shibauchi, and Y. Matsuda, ChemPhysChem {\bf 13},
74 (2012).

\bibitem{25} W. Knafo, S. Raymond, J. Flouquet, B. F{\aa}k, M. A. Adams,
P. Haen, F. Lapierre, S. Yates, and P. Lejay, Phys. Rev. B {\bf
70}, 174401 (2004).

\bibitem {jetp} V. R. Shaginyan, K. G. Popov, and V. A. Khodel, JETP {\bf 116},
848 (2013).

\bibitem{26} D. Shechtman, I. Blech, D. Gratias, and J. Cahn,
Phys. Rev. Lett. {\bf 53}, 1951 (1984).

\bibitem{27} K. Deguchi, S. Matsukawa, N. K. Sato, T. Hattori,
K. Ishida, H. Takakura, and T. Ishimasa, Nature Materials {\bf 11},
1013 (2012).

\bibitem{28} T. Fujiwara, in Physical Properties of Quasicrystals,
(Ed. by Z. M. Stadnik, Springer, 1999).

\bibitem{29} R. Widmer, P. Gr\"oning, M. Feuerbacher, and O.
Gr\"oning, Phys. Rev. B {\bf 79}, 104202 (2009).

\bibitem{30} G. Trambly de Laissardi\`ere, Z. Kristallogr. {\bf
224}, 123 (2009).

\bibitem{pikul} N. Oeschler, S. Hartmann, A. P. Pikul, C. Krellner, C. Geibel, and F.
Steglich, Physica B {\bf 403},  1254 (2008).

\end{thebibliography}
\end{document}